\documentclass[12pt]{article}
\usepackage{ amssymb, amsmath, amsfonts, tikz, tikz-cd, xcolor}

\renewcommand{\a}{{\mathsf a}}
\renewcommand{\b}{{\mathsf b}}
\renewcommand{\c}{{\mathsf c}}
\newcommand{\e}{\epsilon}
\renewcommand{\u}{{\mathsf u}}
\renewcommand{\v}{{\mathsf v}}
\newcommand{\x}{{\mathsf x}}

\newcommand{\A}{{\mathfrak A}}
\newcommand{\AlgProb}{\mbox{\bf AlgProb}}
\newcommand{\B}{{\mathcal B}}

\newcommand{\Cat}{{\mathcal C}}

\newcommand{\cor}{\mbox{Cor}}
\newcommand{\D}{{\mathcal D}}
\newcommand{\dom}{\mbox{dom}}
\newcommand{\mc}{\mathcal}
\newcommand{\mf}{\mathfrak}

\newcommand{\E}{{\mathbb E}}
\newcommand{\Ev}{{\mathcal E}}

\renewcommand{\H}{{\mathbf H}}
\newcommand{\id}{\mbox{id}}
\newcommand{\K}{{\mathbf K}}

\newcommand{\M}{{\mathcal M}}
\newcommand{\N}{{\mathbb N}}
\newcommand{\nat}{\stackrel{\cdot}{\rightarrow}}
\newcommand{\OA}{\mbox{\bf OA}}

\newcommand{\Mod}{\mbox{\bf Prob}}
\newcommand{\PProb}{\mbox{\bf PProb}}
\renewcommand{\P}{{\mathbb P}}
\renewcommand{\Pr}{\mbox{\bf Pr}}
\newcommand{\pr}{\mbox{Pr}}
\newcommand{\Power}{{\mathcal P}}
\newcommand{\Prob}{\Mod}
\newcommand{\R}{{\mathbb R}}
\newcommand{\ran}{\mbox{Ran}}
\renewcommand{\S}{{\mathcal S}}
\newcommand{\Set}{\mbox{\bf Set}}

\newcommand{\supp}{\mbox{supp}}

\newcommand{\V}{{\mathbf V}}
\renewcommand{\tilde}{\widetilde}
\renewcommand{\hat}{\widehat}

\newcommand{\oc}{\mbox{ co }}
\newcommand{\co}{\oc}

\newcommand{\for}[1]{\overrightarrow{#1}}

\newcommand{\tempout}[1]{{}}

\newcounter{thaler}
\newenvironment{mlist}{\begin{list}{\arabic{thaler}}%
{\usecounter{thaler}
\setlength{\rightmargin}{\leftmargin}
\topsep=0pt
\itemsep=0pt
\parskip=0pt
\parsep=0pt
}}{\end{list}}

\usepackage{tikz}
\tikzset{help lines/.style=very thin}

\begin{document} 

\begin{center}{\bf Coarse-graining and compounding as monads}\\
Alex Wilce\\
Susquehanna University,\\
October 10, 2024 

\end{center} 

\begin{abstract} 

Two very basic constructions involving experimental procedures are the formation of coarse-grained versions of experiments, and the formation of branching sequential experiments. The latter allow for the conditioning of states on the results of previous measurements. When one conditions on the results of different coarse-grainings of the same previous experiment, the possibility of interference effects arises.  Here, I show  how to formulate both constructions in terms of monads on a suitable category of (general) probabilistic models. Moreover, I show that these are connected by  distributive law, allowing for a composite monad describing the closure of a probabilistic model under both coarse-graining and sequential measurement. Algebras for all three monads are characterized;  lessons are drawn regarding the possibility of interference and also regarding the formation of sequential products of effects; and connections are made with some themes from the older quantum-logical literature.

\end{abstract}

\section{Introduction}

{The mathematical literature on the foundations of quantum theory have been dominated by the study of three kinds of structures: in both historical order and in order of increasing generality, C*- and von Neumann algebras,  quantum logics (orthomodular posets and lattices with rich state spaces), and most recently, what are often called ``generalized probabilistic theories", or GPTs. In the course of this development, categorical considerations have taken an increasingly prominent place \cite{Abramsky-Coecke, Barnum-Wilce, Shortcut}.


Broadly speaking, a GPT describes physical systems in terms of what may be called 
{\em probabilistic models}: structures specifying a system's states, the possible 
experiments one can perform on the system, and how the former assign probabilities to the 
outcomes of the latter.  In this paper, I consider two basic constructions involving probabilistic models,  one describing the coarse-graining of experiments, the other, the formation of branching compound sequential experiments. The aim here is to recognize both constructions as monads in a suitable category $\Prob$ of probabilistic models,  and to explore some of the implications of this.  Among other things, I will identify the algebras for both monads. 
Subject to a mild additional constraint, algebras for the coarse-graining monad give rise to ``logics" that are orthomodular posets. An algebra for the compounding monad carries an associative sequential product, and its event structure (and, when it exists, its logic) is totally non-atomic. 

I also consider how these two monads interact. A model closed under both sequential composition and coarse-graining will in general support a notion of {\em interference}, of which quantum interference is an instance. As it turns out, the coarse-graining and compounding monads are connected by a distributive law, allowing us to form a composite monad  and corresponding algebras. These latter objects have a very rich and 
interesting structure; strikingly, however, they {\em do not} support interference. Rather, interference arises only when one has a method of coarse-graining that does {\em not} distribute across the sequential product of events. 

The study of sequential composition of experiments has a long history in the older quantum-logical literature; see, e.g., \cite{Foulis, Randall, RF, FR-stability,  Poole}.  In particular, Randall \cite{Randall, RJF} established that quantum logics associated with a kind of free sequential product are non-atomic, and a prescient paper by his student R. Wright \cite{RW-Spin} made explicit the connections among  sequential experiments,  coarse-graining, and interference, in a mathematical framework close to the one used here. Much of what follows can be seen as an updating and further development of ideas originating in this work. 

The audience for this paper likely includes readers who are familiar with the  quantum-foundational literature, and with GPTs in particular, but less so with category-theoretic ideas, and others for whom the reverse is true. I have therefore tried to include enough  detail to make the paper accessible to those with either of these backgrounds. Sections 2 and 3 provide an overview of the general GPT framework, with what I hope is enough exposition to be followable by those coming to this for the first time, but not so much as to become a slog. Categorical ideas are introduced as needed, again in what I hope is enough, but not too much, detail. Sections 4, 5 and 6 contain new material; Section 7 wraps things up with some general conclusions and some specific open questions. As a further aid to readability, I have relegated a significant amount of the more purely technical material to a series of appendices. 

{\bf Acknowledgements} I wish to thank Howard Barnum and Chris Heunen for helpful  comments on earlier drafts of this paper.  Parts of this paper were written while I was a guest at the Perimeter Institute for Theoretical Physics.  I would like to thank my hosts, Lucien Hardy and Rob Spekkens, for their hospitality during that time. Research at Perimeter Institute is supported bythe Government of Canada through the Department of Innovation, Science and Economic Development Canada and by the Province of Ontario through the Ministry of Research, Innovation and Science. 

\section{ (Generalized) Probabilistic Models} 

Different authors mean slightly different things by a ``general(ized) probabilistic theory". The framework sketched here draws on work of Foulis and Randall in the 1970s and 80s (see, e.g., \cite{FR-MMQM, FPR}), enhanced with some category-theoretic ideas.  General references for this section are the survey papers \cite{BW-Foils, Wilce-handbook}. 

A {\bf test space} \cite{FGR-II, Wilce-handbook} is a nonempty collection $\M$ of nonempty sets $E, F, ...$, 
each regarded as the outcome set of some operation, experiment, or {\em test}. The set 
$X := \bigcup \M$ is the {\em outcome-space} of $\M$. A {\em probability weight} on $\M$ 
is a function $\alpha : X \rightarrow [0,1]$ summing to unity on each test. A {\em probabilistic model} --- or 
simply a {\em model}, for short --- is a pair $A = (\M,\Omega)$ where $\M$ is a test space and $\Omega$ is a distinguished set of probability weights on $\M$, called the {\em state space} of the model.  

I write $\M = \M(A)$, $X = X(A)$, and $\Omega = \Omega(A)$ to indicate the test space, outcome-space, and state space of a model $A$. Note that if $\M$ is any test space, we have an associated model $(\M,\Pr(\M))$ where $\Pr(\M)$ is the set of all 
probability weights on $\M$. I will call a model $A$ {\em full} iff it has this form, i.e, $\Omega(A) = \Pr(\M(A))$

{\em Standing Assumption:} To avoid trivialities, I assume that the set $\Omega$ of states is {\em positive}, meaning that for every $x \in X$, there is at least one $\alpha \in \Omega$ with $\alpha(x) > 0$.  It follows  that $\M$ is {\em irredundant}, that is, an anti-chain: if $E, F \in \M$ and $E \subseteq F$, then $E = F$. 

The simplest and most familiar example of a probabilistic model is, of course, a pair $(\{E\},\{p\})$ where $E$ is a single outcome-set and $p$ is a probability weight on $E$. This is the setting for discrete classical probability theory, and is what is usually meant by the phrase ``(discrete) probabilistic model". The generalization represented by probabilistic models as defined above, is a very modest one: we simply allow that there may be more than one experiment, and more than one state, under consideration. Nevertheless, this opens up a very large range of possibilities. The following indicate how this framework accommodates both classical measure-theoretic  and quantum probability theory. 

{\bf Example: Kolmogorov models} Let $(S,\Sigma)$ be a measurable space. The set $\M(S,\Sigma)$ of countable partitions of $S$ by measurable sets, is a test space, 
the outcomes of which are just the non-empty elements of $\Sigma$, and the probability weights on which correspond in an obvious way to probability measures on $(S,\Sigma)$. Accordingly, by a {\em Kolmogorov model} I mean one of the form $(\M(S,\Sigma),\Omega)$ where $\Omega$ is some positive set of probability measures.

{\bf Example: Hilbert models} If $\H$ is a Hilbert space $\H$, let ${\mathcal F}(\H)$, the collection of frames (unordered orthonormal bases) for $\H$, regarded as a test space. The outcome-space, $X(\H)$, is the unit sphere of $\H$. Every density operator $W$ on $\H$ gives rise to a probability weight on ${\mathcal F}(\H)$ defined by $\alpha_{W}(x) = \langle Wx, x \rangle$.\footnote{Gleason's Theorem asserts that if $\dim(\H) > 2$,then every probability weight has this form, but we shall not need this fact.)} By a {\em Hilbert model}, I mean a model $({\mathcal F}(\H),\Omega)$ where $\Omega$ is a positive set of (states associated with) density operators. 

A slight modification of this example, which we might call a {\em projective} Hilbert model, replaces the unit sphere $X(\H)$ with the projective space $\P_{1}(\H)$, that is, the set of rank-one projections on $\H$: tests are {\em projective frames}, maximal pairwise orthogonal sets of rank-one projections; states are still represented by density operators. 

{\bf Example: von Neumann models} Let $\A$ be a von Neumann algebra and let $\P(\A)$ be its projection lattice. The set $\M(\A)$ of countable (resp., finite) partitions of unity in $\P(\A)$ is a test space, with outcome-set $\P(\A) \setminus \{0\}$, and every normal (resp., arbitrary) state $f \in \A^{\ast}$ induces a probability weight by restriction to non-zero projections. By a {\em von Neumann model}, I mean one of the form $(\M(\A),\Omega)$ where $\A$ is a von Neumann algebra and $\Omega$ is a positive set of normal states on $\A$.

{\em Remarks:} 
(1) An {\em orthogonality space} is a pair $(X,\perp)$ where $\perp$ is a symmetric binary relation on $X$. The outcome-space $X = \bigcup \M$ of any test space $\M$ becomes 
an orthogonality space under the relation $x \perp y$ iff $x, y$ are distinct outcomes of some single test.  In all three of the examples above, tests are identical with maximal pairwise orthogonal sets of outcomes. This is a rather special feature. In general, while events are pairwise orthogonal sets of outcomes, a pairwise-orthogonal set of outcomes need not be an event. (For a simple example, consider the test space $\{\{a,x,b\}, \{b,y,c\}, \{c,z,a\}\}$: 
$\{a,b,c\}$ is pairwise orthogonal, but contained in no test.)

(2) A model is {\em locally finite} when all tests $E \in \M(A)$ are finite sets. The examples above are locally finite only when the underlying object is finite in a relevant sense, i.e., the measurable space $S$ is finite, or the Hilbert space $\H$ or von Neumann algebra $\A$ is infinite dimensional. One can also consider locally finite versions of classical and von Neumann models, replacing $\M(A)$ with $\M_o(A)$, the set of finite measurable partitions of $(S,\Sigma)$ or finite partitions of unity in $\P(\A)$. In this case, probability weights correspond to finitely additive probability measures or not-necessarily normal states, respectively; but there is nothing to prevent our requiring $\Omega(A)$ to consist of countably additive probability measures or normal states.

Going forward, where it does not lead to ambiguity it will be convenient to apply any adjective appropriate to either the test space $\M(A)$ or the state space $\Omega(A)$ to the entire model. Thus, for instance, a model $A$ is locally finite iff $\M(A)$ is locally finite, convex iff $\Omega(A)$ is a convex subset of $\R^{X(A)}$, and so on.  

{\bf Events and Perspectivity} An {\em event} for a test space $\M$ is simply an event in the usual probabilistic sense for one of the tests in $\M$; that is, an event is a set $a \subseteq E$ for some $E \in \M$. We write $\Ev(\M)$ for the set of all events of $\M$, and $\Ev(A)$ for $\Ev(\M(A))$. If $\alpha$ is a probability weight on $\M$, we define the probability of an event $a$ in the usual way, that is,  $\alpha(a) = \sum_{x \in a} \alpha(x)$. In particular, $\alpha(E) = 1$ for every test $E \in \M$. Our positivity assumption yields a converse, which we will need 
later:

{\bf Lemma 0:} {\em If $E \in \Ev(A)$ with $\alpha(E) = 1$ for every state 
$\alpha \in \Omega(A)$, then $E \in \M(A)$.} 

{\em Proof:}  Since $E$ is an event, $E \subsetneq F$ for some test $F \in \M$. 
Suppose $y \in F \setminus E$: by positivity, there exists some state $\alpha \in \Omega(A)$ with $\alpha(y) > 0$, whence, $\alpha(F) > 1$, a contradiction. $\Box$ 

We can extend the orthogonality relation from outcomes to events: two events $a, b \in \Ev(\M)$ are {\em orthogonal}, written $a \perp b$, iff they are disjoint and their union is still an event. In this case, $\alpha(a \cup b) = \alpha(a) + \alpha(b)$ for every probability weight $\alpha$ on $\M$.  If $a \perp b$ and $a \cup b \in \M$ --- that is, if $a$ and $b$ partition a test --- then $a$ and $b$ are {\em complements}, and we write $a \co b$.  In general, one event may have many complements. If two events $a, b$ have a common complement --- that is, if there exists some event $c$ with $a \co c \co b$ --- then we say that $a$ and $b$ are {\em perspective}, and write $a \sim b$. 
For convenience, we will often identify an outcome with the corresponding singleton event. Thus,  given $x \in X(A)$ and $a \in \Ev(A)$, we will write, e.g., $x \perp a$ to mean $\{x\} \perp a$, and $x \sim a$ to mean $\{x\} \sim a$. 

Perspectivity plays a fundamental role in the theory of test spaces and probabilistic models. Note that if $a \sim b$, then $\alpha(a) = \alpha(b)$ for all probability weights $\alpha$ on $\M$.  Any two tests are perspective  (both are complementary to $\emptyset$). Conversely,  if $a \in \Ev(A)$ and $a \sim E$ for some $E \in \M(A)$, then $a \in \M(A)$ as well (since $c \oc E$ implies $c = \emptyset$, by irredundance). 


{\bf Example:} Consider the test space $\M(S,\Sigma)$ of 
countable measurable partitions of a measurable space $(S,\Sigma)$. In our usage, an event for this test space is {\em not} an element of $\Sigma$, but a subset of a partition of $S$ by elements of $\Sigma$; that is, an event is a 
pairwise disjoint set $\{a_i\}$ of sets in $\Sigma$.\footnote{Of course, a single element $a \in \Sigma$ can for present purposes be identified 
with the singleton set $\{a\}$, so we may say that our 
use of the term ``event" here is an extension of the 
usual usage.}
Two events $\{a_i\}$ and $\{b_j\}$ are orthogonal 
iff $a := \cup_i a_i$ and $b := \cup_j b_j$ are disjoint (i.e., 
$a_i \cap b_j = \emptyset$ for all $i,j$), complementary 
iff orthogonal and $a \cup b = S$, and, therefore, 
perspective iff $a = b$. Similarly, events in 
$\M(\A)$, where $\A$ is a von Neumann algebra, are pairwise orthogonal sets of non-zero projections; two such sets $\{p_i\}$ and $\{q_j\}$ are orthogonal iff $p := \oplus_i p_i$ and $q := \bigoplus q_j$ are orthogonal (that is, $p_i \perp q_j$ for all $i,j$), complementary iff $b = a'$, and perspective iff $a = b$. 


{\bf Algebraic test spaces and orthoalgebras} 
It is easy to check that if $a \sim b$, then $\alpha(a) = \alpha(b)$ for all probability weights $\alpha$ on $\M$. Also, owing to irredundance, if $a \subseteq b \sim a$, we have $a = b$. A test space $\M$ is {\em algebraic} (or, in older 
terminology, a {\em manual}) \cite{Wilce-handbook} iff, 
for all events $a, b, c \in \Ev(\M)$, if 
$a \sim b$ and $b$ is complementary to $c$, then 
$a$ is also complementary to $c$. If $\M$ is algebraic, 
one can show that  
$\sim$ is an equivalence relation on $\Ev(\M)$, 
with the feature that 
\[a \sim b \perp c \ \Rightarrow a \perp c \ \mbox{and} \ a \cup c \sim b \cup c.\]
This makes it possible to define a partial binary 
operation $\oplus$ on the set $\Pi(\M) := \Ev(\M)/\sim$ of equivalence classes of events by setting 
\[[a] \oplus [b] = [a \cup b] \ \mbox{when} \ a \perp b.\]
It is easy to check that the structure $(\Pi, \oplus)$ 
satisfies the following conditions:  
\begin{mlist} 
\item[(a)] $\oplus$ is associative, in the sense that 
for all $p,q,r \in \Pi$, if $p \perp q$ and $p \oplus q \perp r$ then 
$p \perp (q \oplus r)$ and $(p \oplus q) \oplus r = p \oplus (q \oplus r)$; 
\item[(b)] For every $p \in \Pi$, there exists a unique 
$a' \in \Pi$ with $p \perp p'$ and $p \oplus p' = 1$. 
\item[(c)] If $p \in \Pi$ with $p \perp p$, then $p = 0$. 
\end{mlist} 
For (b), note that if $p = [a]$ where $a \in \Ev(A)$, 
then $p' = [b]$ where $b$ is any event complementary to $a$. 
Abstractly, a structure $(\Pi,\oplus,1)$ satisfying these 
conditions, with $0 := 1'$, is called an {\em orthoalgebra}. 
Thus, for an algebraic test space $\M$, $\Pi(\M)$ is 
an orthoalgebra, called the {\em logic} of $\M$. 
In fact, every orthoalgebra arises as $\Pi(\M)$ for some algebraic test space $\M$. Indeed, if $L$ is an orthoalgebra, let $\D(L)$ denote the set of orthopartitions of its unit: 
that is, a finite set $E \subseteq L \setminus \{0\}$ belongs 
to $\D(L)$ iff $\bigoplus E$ exists and equals $1$.  
This is an algebraic test space, and $L \simeq \Pi(\D(L))$.)  

Any orthoalgebra $L$ is partially ordered by the relation 
\[b \leq a \ \Leftrightarrow \ \exists c \in L ~ b \perp c ~ \& ~ a = b \oplus c.\]
The mapping $a \mapsto a'$ is then an orthocomplementation, $a \perp b$ iff $a \leq b'$, and, in this case, $a \oplus b$ is a minimal (but not necessarily the unique minimal) upper bound for $a$ and $b$.  An orthomodular poset --- an orthoposet $(L,\leq,')$ in which 
$a \leq b \ \Rightarrow  a \vee b$ exists, and $a \leq b' \rightarrow (b \wedge a') \vee a = b$ --- is essentially the same thing as an orthoalgebra in which $a \perp b$ implies 
$a \oplus b = a \vee b$.

\tempout{Historically, the most general algebraic model for an abstract ``quantum logic'' was 
understood to be an {\em orthomodular poset} (OMP):  an 
orthocomplemented poset $(P,\leq,')$ satisfying the conditions 
\begin{itemize} 
\item[(a)] If $a, b \in P$ and $a \leq b'$, then $a \vee b$ exists; 
\item[(b)] If $a, b \in P$ with $a \leq b$, then $(b \wedge a') \vee a = b$. 
\footnote{Note here that by $a \leq b$ implies $b' \leq a'$, so by condition (a), 
$b' \vee a$ exists; hence, so does $(b' \vee a)' = b \wedge a'$.}
\end{itemize} 
It was widely thought that this structure was adequately motivated, e.g., by the axioms 
proposed by Mackey \cite{Mackey}. However,  one can show that an orthomodular poset 
is essentially the same thing as an orthoalgebra in which $a \oplus b = a \vee b$ whenever the former exists. Consequently, an orthomodular lattice is effectively the same thing as a lattice-ordered orthoalgebra. See \cite{Wilce-handbook} for details.  The abundance of 
non-orthocoherent orthoalgebras challenged the widespread assumption that 
}

If $A$ is a probabilistic model, we say that $A$ is algebraic iff $\M(A)$ is algebraic, in which case we write $\Pi(A)$ for $\Pi(\M(A))$, calling this the {\em logic} of $A$.  This is only one of several order-theoretic objects that one can attach to a model $A$.  We now briefly discuss two others. 

{\bf Linearized Models and Effect Algebras}  
If $A$ is a probabilistic model, let $\V(A)$ denote the subspace of $\R^{X(A)}$ spanned by $A$'s state-space, $\Omega(A)$. Ordered pointwise, $\V(A)$ is an ordered vector space with positive cone $\V(A)_+ = \{ \omega \in \V(A) \, | \, \omega(x) \geq 0\}$ for all $x \in X(A)$.  Elements of $\V(A)_+$ are called {\em positive weights} on $A$. Note that every 
$\omega \in \V(A)$ sums to a common value over all tests 
in $\M(A)$. If $\omega \in \V_{+}(A)$ and $\omega \not = 0$, 
$\omega = t\alpha$ for a unique $t > 0$ and $\alpha \in \Omega(A)$. If $t < 1$, we say that $\omega$ is a 
{\em sub-normalized weight} on $\M$.\footnote{Subnormalized weights can be viewed as states for a model arising from $A$ by additing to every test of a common ``null" (or {\em failure}) outcome. The idea of adjoining a null outcome has other uses as well, some of which are discussed in Appendix C.}

The dual space, $\V(A)^{\ast}$, of $\V(A)$, with the 
dual ordering, is an order-unit space with 
unit effect $u$ given by $u_{A}(\alpha) = \sum_{x \in E} \alpha(x)$ for any $\alpha \in \V(A)$ and $E \in \M(A)$ (note that this is independent of $E$).  
A functional $a \in \V(A)^{\ast}$ with $0 \leq a \leq u_{A}$ is called an {\em effect}. Every event $a \in \Ev(A)$ 
gives rise to an effect $\hat{a}$ by evaluation, i.e., 
$\hat{a}(\alpha) = \alpha(a)$,  but in general there will 
be effects not of this form.  

The set $[0,u_{A}]$ of all effects in $\V(A)^{\ast}$ is the most important example of 
what's called an {\em effect algebra}. Abstractly, an effect algebra is a triple 
$(L,\oplus, 1)$ where $\oplus$ is a partial binary 
operation on $L$, $1 \in L$, satisfying (a) and (b) 
in the definition of an orthoalgebra (writing $a \perp b$ 
to mean $a \oplus b$ is defined), and, in place of (c), 
\vspace{-.1in}
\begin{itemize} 
\item[($c'$)] If $a \in L$ with $a \perp 1$, then $a = 0$. 
\end{itemize} 
Note that the structure of the test space $\M(A)$ plays no role in the 
construction of $[0,u_A]$: this depends only on $\Omega(A)$. Thus, 
while not every test space gives rise to an orthoalgebra (only algebraic ones do so), every convex set, and hence, every probabilistic model, gives rise to an effect algebra.   If $A$ does happen to be algebraic, we then have a well-defined effect-algebra embedding $\Pi(A) \rightarrow [0,u_{A}]$ taking $[a] \in \Pi$ to $\hat{a}$. 

\tempout{Say that an element $a$ of an effect algebra $L$ is {\em divisible} iff 
for every $n$, there exists a unique element 
$b$ with $n b := b \oplus \cdots \oplus b$ (with $n$ summands) exists and equals $a$. If $1$ is divisible, say that $L$ is 
divisible. This is the case for $L = [0,u_A]$ for any model $A$. Every effect algebra $L$ gives rise to a test space. Let $\D(L) = \{ E \subseteq L \setminus \{0\} | \bigoplus E = u\}$. If $L$ is divisible, then $\bigcup \D(L) = L \setminus \{0\}$: if $a \in L$ is non-zero and $a \not = a'$, then $\{a,a'\} \in \D(L)$. If $a = a'$, i.e., if $a = \tfrac{1}{2}1$, then $\{a, \tfrac{1}{3}u, \tfrac{1}{6}u\} \in \M(L)$.   

A {\em probability measure} on an effect algebra $L$ is a mapping 
$\mu : L \rightarrow [0,1]$ with $a \perp b \Rightarrow \mu(a \oplus b) = \mu(a) + \mu(b)$ and $\mu(u) = 1$.   Let $\Omega(L)$ denote the convex set of all probability measures 
on $L$. As every probability measure on $L$ defines a probability weight on $\D(L)$, 
we can associate to every divisible effect algebra $L$ a probabilistic model $A(L) := (\D(L),\Omega(L))$. This will satisfy our positivity assumption iff $L$ supports a 
full set of probability weights [check].}

{\bf The Lattice $C(X(A),\perp)$} 
For any set $a \subseteq X(A)$, not necessarily an event, let $a^{\perp} = \{ y \in X | y \perp x \forall x \in a\}$. One says that $a$ is {\em ortho-closed} iff $a = a^{\perp \perp}$. The collection $C(X(A),\perp)$ of ortho-closed subsets of 
$X(A)$, ordered by inclusion, is a complete ortholattice with $\bigvee_i a_i = (\cup_i a_i)^{\perp \perp}$, $\bigwedge_i a_i = \cap_i a_i$, and $a' = a^{\perp}$.   $\M$ is {\em regular} iff $a \sim b \Rightarrow a^{\perp} = b^{\perp}$ for all events $a, b \in \Ev(A)$. If $A$ is both regular and algebraic, we have a natural order-preserving mapping $[a] \mapsto a^{\perp \perp}$ from $\Pi(A)$ to $C(X(A),\perp)$. 
A more detailed account of the relations among $\Ev(A)$, $C(X(A),\perp)$, and (where $A$ is algebraic) $\Pi(A)$ can be found in \cite{Wilce-handbook}.  

{\bf Spin Experiments \emph{\`{a} la} Wright} To conclude this section, we consider a rather concrete class of examples first described by Wright \cite{RW-Spin}, 
from whose paper the following is adapted. These not only illustrate some of the general ideas discussed above, but also bring out the interplay between the {\em coarse-graining} of experiments and the construction of sequential experiments that will concern us below. 

Let $\S_{n}$ denote the set of {\em generalized spin experiments}, 
in which an initial beam of spin-$n$ particles is passed through a SG apparatus, oriented at any angle, and thus split into $2n+1$ outgoing beams. We can then (i) block some or 
all of the outgoing beams; (ii) place a detector in the 
path of one or more beams; (iii) place a {\em detector loop} 
around one or more beams, registering the passage of a 
particle without destroying it.  We can combine two-stage experiments by 
placing further SG magnets in the paths of one or more 
of the (possibly recombined) unblocked outgoing beams. 
By repeating this construction, iterated SG experiments 
having any number of stages can be constructed. 

An ideal SG apparatus for a spin-3/2 system yields four  outgoing beams. To simplify the diagram below, I imagine that one of these beams has been blocked, so that there are three emerging beams. Assuming the system has a non-zero charge, a simple which-path measurement consists of placing detector loops around each of the three paths, 
each wired to an indicator of some kind, giving 
us a test $E$ with three outcomes, say $x, y$ and $z$\footnote{Note that these are arbitrary lables, having nothing to do with spin-directions.} 
We can {\em coarse-grain} $E$ by wiring the loops at $x$ and $y$ (say) to a single display, representing the {\em event} $\{x,y\}$, as in Figure 1 (a).   Alternatively, we can place a single large detector loop (the dashed circle) around the two paths corresponding to $x$ and $y$, wiring this to a single indicator, call it $q$.  
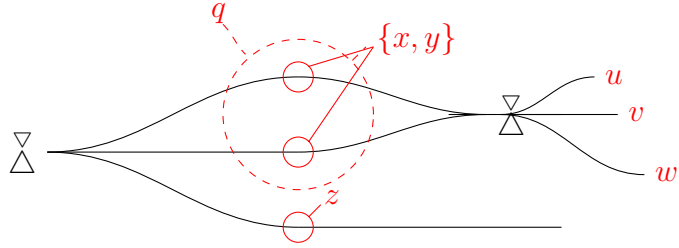
\[
\begin{tikzpicture} 
\node[anchor=west] at (2,0) (splitter) {$\stackrel{\bigtriangledown}{\triangle}$}; 
\node[anchor=west] at (8.5,.5) (splitter2) {$\stackrel{\bigtriangledown}{\triangle}$}; 
\node[anchor=west] at (6,1) (detectorx) {}; 
\node[anchor=west] at (6,0) (detectory) {};
\node[anchor=west] at (6,-1) (detectorz) {};
\node[anchor=west] at (8.5,.5) (combined) {};
\draw (splitter) edge[out=0,in=180] (detectorx);
\draw (splitter) edge[out=0,in=180] (detectory);
\draw (splitter) edge[out=0,in=180] (detectorz);
\draw (6,1) edge[out=0,in=180] (combined);
\draw (6,0) edge[out=0,in=180] (combined);
\draw (6,-1) -- (9.5,-1);
\node at (1,0) (ring) {};
\draw[red] (6,0) circle(.2cm) (6.14,0.14) -- (7,1.4); 
\draw[red] (6,1) circle(.2cm) (6.14,1.14) -- (7,1.4);
\draw[red] (6,-1) circle(.2cm) (6.14,-.86) -- (6.3,-.7);
\node[red, anchor=west] at (6.2,-.6) (outcomez) {$z$};
\node[red, anchor=west] at (6.9,1.5) (outcomexy) {$\{x,y\}$};
\node[red, anchor=west] at (4.7,1.85) (outcomexy) {$q$}; 
\draw[red, dashed] (5,1.7) edge[out=-45,in=135] (5.3,1.3); 
\draw(8,.5) -- (9.5,.5);
\draw[red, dashed] (6,.5) circle (1cm) (6.72,1.22) -- (6.9,1.4);
\node[red] at (10.5,.5) (v) {$v$};
\node[red] at (10.9,-.3) (w) {$w$};
\node[red] at (10.2,1)(u) {$u$};
\draw (8.7,.5) edge[out=0, in=180] (u);
\draw (8.7,.5) edge[out=0, in=180] (v);
\draw (8.7,.5) edge[out=0, in=180] (w);
\end{tikzpicture}
\]
We now have a second test, $F = \{q,z\}$, and a test space 
$\S = \{E,F\}$. The event $\{x,y\}$ and the single outcome $q$ are 
perspective, so they have the same probability, regardless 
of the system's state. However, if we perform a subsequent measurement $G$, say with outcomes $\{u,v,w\}$ 
by placing a second SG apparatus in the 
way of the combined $x, y$ beam, and orienting it at an 
angle $0 < \theta < \pi$ relative to the first apparatus, 
then we find that the conditional probability of, say, 
outcome $u$ for the second measurement depends very much 
on whether or not we performed $E$ or $F$: 
\[\pr(u | \{x,y\}) = \pr(u | x) + \pr(u | y) \not = \pr(u |z).\]
In such a situation, one says that there is {\em interference} between outcomes $x$ and $y$. 

\section{Categories of probabilistic models} 

Probabilistic models can be made into a category in various ways, depending on what kinds of maps one takes as morphisms \cite{FR-MMQM, FPR, Wilce-handbook, BW-Foils}.  The following definition is stricter than some others one could consider, but still general enough for our present purposes. Some other options are discussed and compared in Appendix A. 

{\bf Definition:} Let $A$ and $B$ be probabilistic models. A {\em morphism of } $\phi : A \rightarrow B$ is a mapping $\phi : X(A) \rightarrow X(B)$ such that 
\begin{mlist} 
\item[(a)] For all $x, y \in X(A)$, $x \perp y \Rightarrow \phi(x) \perp \phi(y)$; 
\item[(b)] For each test $E \in \M(A)$, $\phi(E) \in \Ev(B)$;
\item[(c)] $\phi(E) \sim \phi(F)$ for all tests $E, F \in \M(A)$; 
\item[(d)] For each state $\beta \in \Omega(B)$, $\phi^{\ast}(\beta) := \beta \circ \phi$ is a multiple of a state in 
$\Omega(A)$. 
\end{mlist} 

Condition (a) implies that $\phi$ is injective when restricted to tests. We will such a mapping {\em locally injective.} Condition (b) implies that $\phi(a) \in \Ev(B)$ for every event $a \in \Ev(A)$, and it follows that if $a \perp b$ in $\Ev(A)$, then $\phi(a) \perp \phi(b)$ in $\Ev(B)$.  It follows from condition (c) that if $\beta$ is a probability weight on $B$, $\phi^{\ast}(\beta) = \beta \circ \phi$ is a  sub-normalized positive weight on $\M(A)$. Condition 
(d) requires that. if non-zero, this normalize to a state in $\Omega(A)$.

Tests $E, F \in \M(A)$ are perspective by default; condition (c) requires that their images under $\phi$ should still be perspective {\em events} in $\Ev(B)$. In fact, this is enough to guarantee that $\phi$ preserves perspectivity for all 
events: 

{\bf Lemma 1:} {\em Let $\phi : A \rightarrow B$ satisfy 
conditions (a), (b) and (c) above.  Then 
\[a \sim b \ \Rightarrow \ \phi(a) \sim \phi(b)\]
for all events $a, b \in \M(A)$. } 

{\em Proof:}  
If $a \co c$ and $c \co b$, then $\phi(a) \perp \phi(c)$, 
$\phi(b) \perp \phi(c)$, and 
$\phi(a) \cup \phi(c) \sim \phi(c) \cup \phi(b)$. Hence, for some event 
$e \in \Ev(B)$, $\phi(a) \cup \phi(c) \co e$ and 
$\phi(b) \co \phi(c) \co e$, whence, 
$\phi(a) \sim \phi(b)$ with axis $\phi(c) \cup e$. $\Box$

Clearly, the composition of two morphisms 
is again morphism, and the identity mapping on outcomes provides an identity morphism. 
This defines a category $\Prob$ of probabilistic 
models and their morphisms.  A {\em probabilistic theory} can now be defined as a functor 
$\Cat \rightarrow \Prob$, where $\Cat$ is some category of physical systems and physical processes. This point of view 
is already implicit in Examples 1-3 above:  

{\bf Examples:} (a) Let $(S,\Sigma)$ and $(S',\Sigma')$ be measurable spaces, and 
$f : S \rightarrow S'$, a surjective measurable mapping. Then $f^{-1} : \Sigma' \setminus \{\emptyset\} \rightarrow \Sigma \setminus \{\emptyset\}$ defines an injective morphism $\M(S',\Sigma') \rightarrow 
\M(S,\Sigma)$. If $\mu$ is a probability measure on 
$\mu \circ \phi^{-1}$ is a probability measure on $\Sigma'$, 
and we have a contravariant functor 
$\Cat \rightarrow \Prob$ where $\Cat$ is the category 
of measure spaces and measurable surjections. 

(b) Let $V : \H \rightarrow \K$ be an isometry from a Hilbert space $\H$ to a Hilbert space $\K$. Restricted to the unit sphere of $\H$, this defines an injective morphism from the Hilbert model associated with $\H$ to that associated with $\K$. This gives us a covariant functor from the category of Hilbert spaces and isometries to $\Prob$. 

(c) Similarly, any injective $\ast$-homomorphism or $\ast$-antihomomorphism from one von Neumann algebra to another induces a morphism between the associated von Neumann models. 

As these examples illustrate, our concept of morphism is somewhat restrictive. In order to capture, e.g., arbitrary measurable mappings, partial isometries, or arbitrary $\ast$-homomorphisms between von Neumann algebras, one needs to consider {\em partial} morphisms. These are discussed in Appendix A. 
However, in the body of this paper, I will consider only totally defined morphisms.

\noindent{\bf Definition:} A morphism $\phi$ (whether 
of test spaces or of models) is  
\begin{mlist}
\item[(i)] {\em test-preserving} iff it maps 
tests to tests; 
\item[(ii)] an {\em embedding} iff it is test-preserving and 
(gloabally) injective. 
\end{mlist} 
Test-preserving morphisms are particularly well-behaved. In particular, if $\phi$ is test-preserving, $\phi^{\ast}(\beta) \in \Omega(A)$ for 
every $\alpha(\beta) \in \Omega(B)$. I will denote the category of probabilistic models and test-preserving morphisms by $\Prob_1$. 

A test-preserving mapping $X(A) \rightarrow X(B)$ satisfying conditions (a)  in the definition of a morphism automatically satisfies conditions (b) and (c) as well.  In other words, a test-preserving morphism is a locally-injective, test-preserving mapping. Another observation that is sometimes useful is that 
if a morphism $\phi : X(A) \rightarrow X(B)$ preserves 
one test --- say, $\phi(E) \in \M(B)$ for a particular 
$E \in \M(A)$ ---  
then $\phi$ preserves all tests. For if $F \in \M(A)$, 
we have $F \sim E$, whence $\phi(F) \sim \phi(E)$, 
whence, $\phi(F) \in \M(A)$.   

Looking back at the examples above, we see that the morphisms associated with measurable 
surjections are always embeddings. Those associated with isometries are test-preserving only for unitary mappings, while those associated with injective $\ast$-homomorphisms of von Neumann algebras are  test-preserving iff the homomorphism is unital, in which case, 
they are embeddings. For instance, if $\mf{A}$ and $\mf{B}$ are von Neumann algebras, the mapping $\phi : a \mapsto a \otimes 1_{\mc{B}}$ gives rise to an embedding of the von Neumann model corresponding to $\mf{A}$ into that corresponding to $\mf{A} \otimes \mf{B}$. 

One simple example of an embedding arises as follows: let ${\mathcal A}$ be any collection of mutually perspective events in $\Ev(A)$. Define a model $(\M,\Omega_{\M})$, where $\Omega_{\M}$ consists of all normalilzed versions of restrictions to $\bigcup \M$ of states in $\Omega(A)$. The inclusion mapping $\bigcup {\mathcal A} \rightarrow X(A)$ is then an embedding. 

The following shows that left inverses of embeddings are test-preserving.

{\bf Lemma 2:} {\em Let $\phi : A \rightarrow B$ be an 	 embedding, and let $\psi : B \rightarrow A$ 
be a morphism with $\psi \circ \phi = \id_{B}$. Then $\psi$ is test-preserving.} 

{\em Proof:} Let $E \in \M(A)$. Since $\phi(E)$ is an event of $B$, we can find some $F \in \M(B)$ with 
$\phi(E) \subseteq F \in \M(B)$, then $E = \psi(\phi(E)) \subseteq \psi(F) \in \Ev(A)$. Hence, by irredundance of $\M(A)$, $E = \psi(F)$. Now for any other test $F' \in \M(B)$, $F' \sim F$, 
so $\psi(F') \sim \psi(F) = E$, whence, $\psi(F') \in \M(A)$ as well.  $\Box$

It is also straightforward that if $\phi : A \rightarrow B$ and $\psi : B \rightarrow A$ 
are mutually inverse morphisms, then $\phi : X(A) \rightarrow X(B)$ is a bijection, $\psi$ is its inverse, 
and both $\phi$ and $\psi$ are test-preserving. 
In other words, an {\em isomorphism} of models $A$ and $B$ 
is just what we'd expect: a bijection $X(A) \rightarrow X(B)$ preserving tests, and states, in both directions.

{\em Remark:} Of course, $\Prob$ has a great many subcategories of interest, including the category 
$\Prob_1$ of models and test-preserving morphisms, 
and the category $\AlgProb$ of algebraic 
models and (arbitrary) morphisms. If $\Pi$ and $\Pi'$ are orthoalgebras, an 
{\em orthomorphism} $f : \Pi \rightarrow \Pi'$ is a mapping that pereserves $\perp$ and the partial operation $\oplus$. 
These evidently compose, providing us with a category 
category, $\OA$, of orthoalgebras and orthomorphisms. 
It is straightforward that any morphism 
$\phi : A \rightarrow B$ between algebraic models 
gives rise to an orthomorphism $\hat{\phi} : \Pi(A) \rightarrow \Pi(B)$ between their logics, given by 
$\hat{\phi}([a]) = [\phi(a)]$ for any event $a \in \Ev(A)$. 
Thus, we have a functor $\Pi : \AlgProb \rightarrow \OA$.

\section{The Coarsening Monad} 

It is always possible to ``coarse-grain" a single experiment by partitioning its outcome-set $E$ into a number of events $\{a_i\}$: when an particular outcome in $x \in E$  obtained, one records only the occurence of the event $a_i$ with 
$x \in a_i$. 

{\bf Definition:} The {\bf coarsening} of a test space $\M$ is the test space $\M^{\#}$ consisting of all partitions of tests in $\M$.  

Thus, a test in $\M^{\#}$ is a pairwise-disjoint family 
$\{a_i\}$ of non-empty events in $\Ev(\M)$ with $\bigcup_i a_i \in \M$.  The outcome-set of $\M^{\#}$ is $\Ev(\M) \setminus \{\emptyset\}$. Every probability weight on 
$\M$ lifts to a probability weight on $\M^{\#}$ given by $\alpha(a) = \sum_{x \in a} \alpha(x)$, and every probability 
weight on $\M^{\#}$ has this form for a probability on $\M$.
 The coarsening of a model $A$ is the model $A^{\#}$ where 
$\M(A^{\#}) = \M(A)^{\#}$, and $\Omega(A^{\#})$ consists of all lifts of states in $\Omega(A)$ to probability weights 
on $\M^{\#}(A)$.

The following is straightforward:

{\bf Lemma 3:} {\em If $A$ is algebraic, so is $A^{\#}$, 
and in this case there is an 
isomorphism 
$\phi : \Pi(A) \simeq \Pi(A^{\#})$ given by  
$\phi([a]) = [\{a\}]$ where $[a] \in \Pi(A)$ is the 
perspectivity class of $a$ in $\Ev(A)$, and $[\{a\}] \in \Pi(A^{\#})$ is the perspectivity class of $\{a\}$ in $\Ev(A^{\#})$. 
}

{\bf Lemma 4:} {\em Let $\phi : A^{\#} \rightarrow A$ be a 
morphism such that $\phi(\{x\}) = x$ for every $x \in X(A)$.  Then 
\begin{mlist} 
\item[(a)] $\phi$ is test-preserving 
\item[(b)] For all $a \subseteq E \in \M(A)$, 
$(E \setminus a) \cup \{\phi(a)\} \in \M(A)$. In 
particular, $\{\phi(a)\} \sim a$. 
\end{mlist} }

{\em Proof:} (a) follows directly from Lemma 2. 
For part (b), let 
\[E' = \{ a\} \cup \{\{x\} | x \in E \setminus a\} \in \M(A^{\#}),\] 
and use the fact that $\phi$ is test-preserving. $\Box$ 

Recall that a {\em monad} in a category $\Cat$ is a 
functor $T : \Cat \rightarrow \Cat$, together with 
a pair of natural transformations $\eta : \id \rightarrow T$ 
and $\mu : T^2 \rightarrow T$ (called the {\em unit} and 
{\em multiplication} of the monad) such that the 
following commute for all objects $A \in \Cat$: 
\[
\begin{tikzcd}
T^3(A)  \arrow[d, "\mu_{T(A)}" left] \arrow[rr, "T(\mu_{A})"] & & T^2 A \arrow[d, "\mu_A"] 
\arrow[d, "\mu_A"] \\
T^2(A)  \arrow[rr, "\mu_{A}"] & & T(A)\\
\end{tikzcd}
\hspace{.2in} 
\begin{tikzcd}
T(A) \arrow[drr, "\id_{A}" below] \arrow[rr, "T(\eta_A)"] & & T^2(A) \arrow[d, "\mu_A"] & & T(A) \arrow[ll, "\eta_{T(A)}" above] \arrow[dll, "\id_{A}"]\\
 & & T(A) & & \\
 & & & & 
\end{tikzcd}
\]
One can think of a monad as a generalized ``closure" operation. Indeed, if $\Cat$ is a posetal category, a monad is precisely a closure in the order-theoretic sense. 
Some basic examples in $\Set$ are the covariant power-set 
functor and the functor taking a set to the free semigroup 
it generates, regarded as a set. (The reader who has 
not encountered this idea before will find it an excellent exercise to determine the appropriate unit and counit, and to verify the two conditions above, in each of these cases.)

Any morphism $\phi : A \rightarrow B$ in $\Prob$ yields a morphism  
$\phi^{\#} : a \mapsto \phi(a) = \{\phi(x) | x \in a\}$ from $A^{\#}$ to $B^{\#}$, giving us an endofunctor $\#$ on $\Mod$. In fact, it is a monad. 
The unit and multiplication are 
given by the morphisms 
\[\eta_A : A \rightarrow A^{\#}, \ \eta_A(x) = \{x\}\]
and 
\[\mu_{A} : A^{\# \#} \rightarrow A^{\#}, \ \mu_A (a) = \bigcup a\]
for any $x \in X(A)$ and $a \in \Ev(A^{\#\#})$. 
That these satisfy the necessary identities is verified in Appendix D.

{\bf $\#$-Algebras, Coherences, and Cohesions}  
Recall that an {\em algebra} for a monad $T$ --- more briefly, a {\em $T$-algebra} --- is an object $A$ plus a morphism 
$\phi : T(A) \rightarrow A$ such that the following diagrams commute:
\[
\begin{array}{ccc}
\begin{tikzcd}
T^2(A) \arrow[d, "\mu_{A}" left] \arrow[rr, "T(\phi)"] 
& &  
T(A)\arrow[d, "\phi"] \\
T(A) \arrow[rr, "\phi"] & & A\\
& \mbox{(I)} &
\end{tikzcd}
& 
\mbox{and}                                                                                                                                                                                                                                                                                                                &                                                                                                                                                                                                                                                                                                                                                                                                                                                                                                                                                                                                                                
\begin{tikzcd}
A \arrow[drr, "\id_{A}" below] \arrow[rr, "\eta_{A}"] & & T(A)\arrow[d, "\phi"] \\
    & & A \\
    & \mbox{(II)} & 
\end{tikzcd}
\end{array} 
\]

Unpacking this in the case where $T$ is the coarsening monad 
on on $\Prob$, we find that 
a morphism 
$\sigma : A^{\#} \rightarrow A$ turns $A$ into a $\#$-algebra iff
\begin{itemize} 
\item[(i)] $\sigma(\bigcup_{i} a_i) = \sigma\{\sigma(a_i)\}$ for jointly orthogonal sets $\{a_i\}$ of events; 
\item[(ii)] $\sigma(\{x\}) = x$ 
for all $x \in X(A)$.  
\end{itemize} 
Call such a morphism a {\em coherence} on $A$.  
A $\#$-algebra, then, is a model with a designated coherence.  

{\bf Examples} A classical model $(\M(S,\Sigma),\Omega)$ associated with a measurable space has a coherence given by 
$\sigma(\{a_i\}) = \bigcup_i a_i$ where $\{a_i\}$ is any 
pairwise-disjoint collection of nonempty sets $a_i \in \Sigma$. A von Neumann model $(\M(\A),\Omega)$ has a coherence 
given by $\sigma(\{p_i\}) = \bigoplus p_i$ where 
$\{p_i\}$ is any pairwise orthogonal family of projections 
in $\A$.  

{\em Remark:} The term {\em coherence} is motivated by quantum mechanics. If $\{p_i\}$ are mutually orthogonal  projections, say in a von Neumann algebra $\A$, one calls $\oplus_i p_i$ is a {\em coherent} combination of the projections $p \in a$, while the event $\{p_i\}$ is an {\em incoherent} coarsening.  As we saw in the example of iterated spin experiments, the distinction is crucial in sequential experiments, where interference effects only manifest themselves when one conditions on coherent combinations. 

By Lemma 4, a coherence $\sigma$ is always test-preserving, 
and satisfies $\sigma(a) \sim a$ for every $a \in \Ev(A)$. 
It follows that if $\M(A)$ is algebraic, $\sigma$ 
also satisfies 
\begin{equation}
\sigma(a) \in E \Rightarrow (E \setminus \{\sigma(a)\}) \cup a \in \M(A). 
\end{equation} 

{\bf Definition:} A coherence $\sigma$ on $A$ satisfying (\theequation) is a {\em cohesion}. A $\#$-algebra 
$(A,\sigma)$ is {\em cohesive} iff $\sigma$ is a cohesion. 

Thus, if $A$ is algebraic, any coherence on $A$ is a cohesion. 
As we will now see, the existence of a cohesion can be a strong constraint. 

{\bf Cohesions and quantum logics} During the 1960s and `70s, the most general algebraic model of an abstract ``quantum logic'' was understood to be an orthomodular poset. As we'll now see, this structure arises naturally from the existence of a sufficiently well-behaved coarsening. Call a test space {\em projective} 
iff \[x \sim y \Rightarrow x = y\] 
for all outcomes $x, y \in X = \bigcup \M$.  A model 
$A$ is projective iff its test space $\M(A)$ is projective. 
 
The test spaces $M(S,\Sigma)$ and $\M({\mathfrak A})$ associated with a measurable 
space or a von Neumann algebra are projective. The frame-test space ${\mathcal F}(\H)$ of a Hilbert space is not projective (outcomes are 
unit vectors, and are perspective iff they differ by a complex factor of modulus $1$), but the corresponding projective frame manual ${\mathcal F}_{\P}(\H)$ is --- hence our terminology.  More generally, If $\M$ is algebraic, we can ``projectivize" it by considering its image in $\D(\Pi(\M))$ under the mapping $x \mapsto [x]$, that is, the test space $\M_{\Pi} := \{ \{ [x] | x \in E\} | E \in \M\}$. The test space $\D(\Pi(\M))$ itself is projective.  However, in general, $A^{\#}$ is not projective, even if $A$ is. 

{\bf Lemma 5:} {\em Let $(A,\sigma)$ be a $\#$-algebra. The 
following are equivalent:
\begin{mlist} 
\item[(a)] $A$ is cohesive and projective 
\item[(b)] $A$ is cohesive and 
\begin{equation} 
a \sim b \Rightarrow \sigma(a) = \sigma(b)\end{equation} 
\item[(c)] $A$ is algebraic and projective 
\end{mlist} }

{\em Proof:} The only nontrivial part 
is (b) $\Rightarrow$ (c). If $a,b,d \in \Ev(A)$ 
with $a \sim b$ and $b \co d$, then 
$\sigma(a) = \sigma(b)$ by (\theequation). 
Since any coherence is test preserving, 
$\sigma(b) \co \sigma(d)$, whence, 
$\sigma(a) \co \sigma(d)$, whence 
$\{\sigma(a),\sigma(d)\} \in \M(A)$. Since 
$a \sim \sigma(a)$ and $\sigma$ is a cohesion, 
$a \cup \{\sigma(d)\} \in \M(A)$. Since 
$\{\sigma(d)\} \sim d$, the fact that 
$\sigma$ is a cohesion also gives us  
$a \cup d \in \M(A)$. Thus, $\M(A)$ is algebraic. 
To see that it's projective, observe that if 
$x, y \in X(A)$ with 
$\{x\} \sim \{y\}$, then $x = \sigma(\{x\}) = \sigma(\{y\}) = y$. $\Box$. 

Two further conditions on a test space that are important in connecting test spaces to orthomodular structures are {\em coherence} and {\em regularity} \cite{FPR, Wilce-handbook}
 
{\bf Definition:} If $\M$ is a test space and 
$a \subseteq \bigcup \M$, let $a^{\perp}$ denote the set of outcomes orthgonal 
to every outcome in $a$. $\M$ is said to be 
\begin{mlist} 
\item[(i)] 
{\em coherent} iff, for all events $a, b \in\Ev(\M)$, 
$a \subseteq b^{\perp}$ implies $a \perp b$, and 
\item[(ii)] {\em regular} iff $a \sim b$ implies $a^{\perp} = b^{\perp}$. \footnote{Yes, yes: we now have to speak of, e.g., coherences on coherent models. I'm not sure what to do about this. Suggestions for a less awkward term for  a $\#$-structure are welcome.} 
\end{mlist} 

We will need the following standard results: 

{\bf Lemma 6} 
{\em \begin{mlist} 
\item[(a)] A coherent test space is regular iff it is algebraic; 
\item[(b)] The logic of a coherent algebraic test space 
is an orthomodular poset (OMP). 
\end{mlist} 
}

{\em Proof:} For (a), see \cite[Lemma 113]{Wilce-handbook} and for (b), see \cite[Lemma 4.1]{Wilce-TSOA} $\Box$

{\bf Definition:} A model $A$ is {\em unital} if, for ever $x \in X(A)$, there exists 
a state $\alpha \in \Omega(A)$ with $\alpha(x) = 1$, and {\em strongly unital} if 
for every pair of distinct outcomes $x, y \in X(A)$ with $x \not \perp y$, there exits 
a state $\alpha \in \Omega(A)$ with $\alpha(x) = 1$ and $\alpha(y) > 0$. 

{\em Remark:} If $(L,\leq, ')$ is an orthomodular poset, 
then a set $\Omega$ of probability measures on $L$ is 
{\em strong} iff $\alpha(p) = 1 \Rightarrow \alpha(q) = 0$ 
for all $\alpha \in \Omega$ implies that $p \perp q$. 
In parts of the older literature, a {\em quantum logic} 
was defined as a pair $(L,\Omega)$ where $L$ is an OMP 
and $\Omega$ is a strong set of probability measures thereon. 
Evidently, the model $(\D(L),\Omega)$ is strongly unital iff 
$\Omega$ is strong.

{\bf Lemma 7:} {\em Let $A$ be strongly unital. Then $A$ is regular.} 

{\em Proof:} Let $a \sim b$ in $\Ev(A)$. It is enough to show that $b^{\perp} \subseteq a^{\perp}$. If 
$x \not \in a^{\perp}$, then we can find $z \in a$ and $\alpha \in \Omega(A)$ with $\alpha(x) = 1$ and 
$\alpha(z) > 0$. Since $a \sim b$, $\alpha(b) = \alpha(a) > 0$, so there must exist some $z' \in b$ with $\alpha(z') > 0$, whence, $x \not \perp z'$, and so $x \not \in b^{\perp}$. $\Box$ 

The following is related to Theorem 117 in \cite{Wilce-handbook}. (See also \cite{Gudder} for similar results.)

{\bf Proposition 1:} {\em Let $A$ be regular.  If $A$ has a cohesion, then it is coherent and algebraic, and hence $\Pi(A)$ is an OMP.} 

{\em Proof:}  Let $\sigma$ be a cohesion on $A$. By the previous Lemma, $A$ is regular, so if we can show it is coherent, the rest follows from Lemma 7. Let $a, b \in \Ev(A)$ with $a \subseteq b^{\perp}$. By Lemma 4, $\sigma(a) \sim a$, so regularity gives us $\sigma(a) \in b^{\perp}$. Since $b \sim \sigma(b)$, another application of regularity gives us $\sigma(a) \in \sigma(b)^{\perp}$, i.e., $\sigma(a) \perp \sigma(b)$. Thus, there exists a test $E \in \M(A)$ with $\sigma(a), \sigma(b) \in E$. Since $\sigma$ is a cohesion, $E' := E \setminus \sigma(b) \cup b \in \M(A)$. Since $\sigma(a) \not = \sigma(b)$, $\sigma(a) \in E'$, and the definition of a cohesion yields a test 
\[E'' := E' \setminus \sigma(a) \cup a \in \M(A)\]
We now have $a, b \subseteq E''$; since $a \subseteq b^{\perp}$, $a \cap b = \emptyset$, and it follows that $a \perp b$. $\Box$ 

It follows that a strongly unital model having a cohesion is regular, algebraic, and coherent.  Since a coherence on an algebraic model is automatically 
a cohesion, we also have 

{\bf Corollary 1:} {\em Any regular (in particular, a strongly unital) algebraic $\#$-algebra\footnote{Again, the terminology is very awkward. In this case, 
both the adjective {\em algebraic} and the term {\em $\#$-algebra} are dictated by traditional usage. It would help here to have an alternative name for a 
$\#$-algebra, but I have not been able to think of one. Suggestions would be again be welcome!} is coherent, and its logic is an OMP.}

It follows that the logic of a strongly unital and cohesive, or, equivalently, a strongly unital and algebraic, $\#$-algebra is an OMP.  In view of Lemma 6, we see that locally finite, cohesive, strongly unital models essentially {\em are} 
OMPs equipped with a strong set of states.  

{\em Remark:} One can show that an orthoalgebra $L$ is an OMP iff pairwise 
orthogonal triples $a,b,c \in L$ are jointly orthogonal, meaning that $a \perp (b \oplus c)$. 
This condition, variously called {\em ortho-coherence} and {\em Specker's Principle}, is 
implied by, but weaker than, the coherence of $\M$ with $L = \Pi(\M)$. It has widely been seen as unmotivated, and there is a small literature 
(see, e.g., \cite{Specker, Kunjwal-Heunen-Fritz, Bacciagaluppi} devoted to 
providing it with some motivation. The conditions 
given above all have a straightforward operational interpretation, and thus, provide an alternative ``derivation" of a strong form of Specker's principle.

\section{The Compounding Monad}

We now turn to the second of our two basic constructions: the closure of a test space or model under the formation of branching sequential 
{\em compound} 
experiments. 

{\bf Two-stage tests and the forward product}  A simple model for a two-stage compound experiment, possibly 
involving two different models, is as follows. Let $E \in \M(A)$; for each outcome $x \in E$, select 
a test $F_{x} \in \M(B)$. Perform the test $E$, and upon obtaining outcome $x$, 
perform the pre-selected test $F_{x}$. If this yields outcome $y$, record the pair $(x,y)$ as the outcome of the two-stage test. The outcome set for 
this experiment is then 
\[\bigcup_{x \in E} \{x\} \times F_{x}.\]  

We can now define a model $\for{AB}$, the {\em forward product}, of $A$ and $B$ \cite{FR-TP}, as follows: $\M(\for{AB})$ is the collection of all such two-stage tests. 
Thus, $X(\for{AB}) = X(A) \times X(B)$, and 
$(x,y) \perp (u,v)$ in $X(\for{AB})$ iff $x \perp u$ or $x = u$ and $y \perp v$. Probability weights on $\M(\for{AB})$ are uniquely defined 
by pairs $(\alpha, \beta)$ where $\alpha \in \Pr(\M(A))$ 
and $\beta \in \Pr(\M(B))^{X(A)}$ via the formula 
\[(\alpha;\beta)(x,y) := \alpha(x)\beta_{x}(y).\]
We take $\Omega(\for{AB})$ to consist of those weights 
$(\alpha;\beta)$ with $\alpha \in \Omega(A)$ and 
$\beta_{x} \in \Omega(B)$ for all $x \in X(A)$. 
In the special case in which $\beta \in \Omega(B)^{X}$ is constant, 
i.e., $\beta_{x} = \beta \in \Omega(A)$ for every $x \in X(A)$, 
then $(\alpha;\beta)(x,y) = \alpha(x)\beta(y)$. 
We will denote this state as $\alpha \otimes \beta$. 

{\bf Lemma 8:} {\em If $A$ and $B$ are both algebraic, 
both coherent, both regular, or both strongly unital, 
then so is $\for{AB}$. } 

{\em Proof:} The proof that $\for{AB}$ is algebraic if 
$A$ and $B$ are algebraic can be found in \cite[Theorem 3.2.2]{PLock}, and the remaining parts are fairly straightforward. See Appendix C for the details. $\Box$ 

{\bf Marginal and conditional states} If $\omega = (\alpha;\beta)$, then 
$\alpha(x) = \omega(xF)$ for any $F \in \M(B)$, and 
$\beta_{2|x}(y) = \omega(x,y)/\omega(xF)$; that is, 
$\alpha$ is $\omega$'s {\em marginal} state of $\omega$ on $A$ 
and $\beta = \omega_{2|x}$ is $\omega$'s {\em conditional} 
state on $B$. This gives us 
a diachronic version of the law of total probability, 
\[\omega(x,y) = \sum_{x \in E} \omega_{1}(x)\omega_{2|x}.\]
Notice, however, that this is strictly one-way: 
if we attempt to define a marginal on $B$ and conditional 
states on $A$ --- that is, if we want to {\em retrodict} --- 
we find that 
\[\omega_{2,E}(y) := \sum_{x \in E} \omega(x,y) = 
\sum_{x \in E} \alpha(x) \beta_{x}(y)\]
depends essentially on the choice of the initial test $E$; 
hence, so does the conditional state 
\[\omega_{1|E, y}(x) = \frac{\omega(x,y)}{\omega_{1, E}(x)}.\]

{\bf The compounding of a model} One can enlarge a model $A$ to obtain a model, $A^{c}$, the {\em compounding} of $A$ \cite{FR-stability, FR-conditioning} that is closed under the formation of compound tests having any number of stages --- or, 
to put it another way, such that $\for{A^{c} A} \simeq A^{c}$. The outcome-set $X(A^c)$ is the free monoid $X(A)^{\ast}$ on $X(A)$ with identity element $\epsilon$ (the empty string). We identify $X(A)$ with the subset of $X(A)^{\ast}$ consisting of length-one strings. $\M(A^{c})$ is the smallest collection $\D$ of subsets of $X(A)^{\ast}$ containing 
$\{\epsilon\}$ and closed under the formation of sets of the form 
$\bigcup_{x \in E} x F_{x}$ 
where $E \in \D$ and, for every $x \in E$, $F_{x} \in \M(A)$. By construction, 
$\M(A) \subseteq \M(A^{c})$. Every set $E \in \M(A^{c})$ consists of reduced strings of some bounded length 
(since sets in $\epsilon$ has length $0$ and the collection of sets bounded in this way  has the required closure property). 

To complete the description of the model $A^{c}$, we need 
to specify the states. Probability 
weights on $\M(A^{c})$ are associated with 
{\em transition functions} $f : X^{\ast} \times X \rightarrow [0,1]$ with $f(\a, \cdot) \in \Pr(\M(A))$ for all $\a \in X^{\ast}$. Given this data, one defines a probability weight $\omega_{f}$ recursively by 
$\omega_{f}(\epsilon) = 1$ and  $\omega_{f}(\a x) = \omega_{f}(\a) f(\a,y)$. We say that $\omega_{f}$ is a {\em state} of 
$A^{c}$ iff $f(\a, ~\cdot~ ) \in \Omega(A)$ for every $\a \in X^{\ast}$.  

The following is proved in Appendix C:

{\bf Proposition 2:} {\em If $A$ is unital, strongly unital, regular, 
or algebraic, then so is $A^{c}$. If $A$ is coherent and locally finite, 
then $A^{c}$ is also coherent.} 

Given morphisms $\phi : A \rightarrow A'$ and, for 
every $x \in X(A)$, morphisms $\psi_{x} : B \rightarrow B'$, one might hope to define a morphism 
$(\phi;\psi) : \for{AB} \rightarrow \for{A'B'}$ by 
setting 
\begin{equation} 
(\phi;\psi)(x,y) = (\phi(x),\psi_{x}(y)) 
\end{equation} 
for all 
$(x,y) \in X(\for{AB})$. However, in almost all situations, in order for this mapping to preserve perspectivity of events, $\psi_{x}$ must be test-preserving for all $x \in X(A)$. For the full story, see Appendix B.  For now, we only note that if $\psi$ {\em is} test-preserving, then (\theequation) {\em does} define a morphism. Again, see Appendix B for this. 

Any function $\phi : X(A) \rightarrow X(B)$ extends uniquely 
to a function $X(A)^{\ast} \rightarrow X(B)^{\ast}$. If $\phi$ is a test-preserving morphism 
from $A$ to $B$, then the remarks above, plus a simple inductive argument, show this extension defines a test-preserving morphism $\phi^{c} : A^{c} \rightarrow B^{c}$. This is spelled out in Appendix C. 

Thus, $(~\cdot~)^{c}$ is an endofunctor on $\Prob_{1}$. 
The natural semigroup homomorphisms 
$X^{\ast \ast} \rightarrow X^{\ast}$ (concatenation) and $X \subseteq X^{\ast}$ define morphisms in $\Prob_{1}$, and, 
since they also make $(~\cdot~)^{\ast}$ a monad in $\Set$, 
we see that $(~\cdot~)^{c}$ is a monad in $\Prob_{1}$, which 
we'll call the {\em compounding monad}.

{\bf Sequential Models} Algebras for the compounding monad are easily characterized. 

{\bf Definition:} Let $X$ be a monoid. A (non-empty) family of non-empty sets $\M \subseteq 2^{X}$ is 
{\em inductive} iff, for every $E \in \M$ and 
every function $F: E \rightarrow \M$, the set 
\[\bigcup_{x \in E} x F_{x} = \{ xy | x \in E, y \in F_x\}\]
belongs to $\M$.  A {\em sequential test space} is a test space $\M$ such that $X = \bigcup \M$ is a monoid, $\M$ is inductive, and for all $x,y,z,w \in X(A)$, 
\begin{equation} 
x \perp y \ \Rightarrow zx \perp zy  \ \mbox{and} \  xz \perp yw.
\end{equation} 
A {\em sequential product} on $\M(A)$ is any associative, 
unital binary operation making $\M(A)$ a sequential test space. 

Note that by (\theequation), left multiplication by a fixed 
$x \in X(A)$ gives us a $\perp$-preserving mapping 
$X(A) \rightarrow X(A)$, and since $\M(A)$ is 
inductive, $xE \sim xF$ for any 
$E, F \in \M(A)$. Thus, left multiplication by $x$ is a morphism of test spaces.

There is another way to put all this.  Since 
$X(\for{AA}) = X(A)\times X(A)$, any binary operation 
$\pi : X(A) \times X(A) \rightarrow X(A)$ can equally 
well be regarded as a mapping $X(\for{AA}) \rightarrow X(A)$. If $\pi$ is a sequential product on a 
test space $\M(A)$, then the fact that $\M(A)$ is inductive amounts to the assertion that $\pi$ takes $\M(\for{AA})$ into $\M(A)$, and condition (\theequation) makes it injective on tests. Thus, $\pi$ defines a test-preserving morphism of test spaces $\pi : \M(\for{AA}) \rightarrow \M(A)$, and it is easy to see that, conversely, any such morphism will be a sequential product, proivded that, understood as a binary operation on $X(A)$, it is associative and has a unit. This suggests a natural definition of a sequential 
probabilistic model: 

{\bf Definition:} A {\em sequential product} on 
a model $A$ is an associative, unital binary operation 
$\pi : X(A) \times X(A) \rightarrow X(A)$ that is 
also a test-preserving morphism $\for{AA} \rightarrow A$. 
{\em sequential model} is a probabilistic model equipped with a designated sequential product. 

In more down-to-earth terms, then, a sequential model 
is a model $A$ such that $\M(A)$ is a sequential test space, 
and, where $\pi$ is the monoidal product on $X(A)$, 
$\pi^{\ast}(\omega) \in \Omega(\for{AA})$ for all 
$\omega \in \Omega(A)$. This, in turn, simply means that 
\[\omega_1(x) := \sum_{y \in F} \omega(xy)\]
and 
\[\omega_{x}(y) := \frac{\omega(xy)}{\omega_1(x)}\]
both belong to $\Omega(A)$.


It follows that if $(A,\pi)$ is a sequential model, every state $\omega \in \Omega(A)$ has marginal and conditional states, defined for $x, y \in X(A)$ by $\omega_1 = \pi^{\ast}(\omega)_1$ and $\omega_{x} = \pi^{\ast}(\omega)_{2|x}$, assuming in the latter case that $\omega(x) \not = 0$. More concretely,  if $x, y \in X(A)$ we have for $x \in X(A)$ and any $F \in \M(A)$, 
\[\omega_1(x) = \sum_{y \in F} \omega(xy) \ \mbox{and, for $\omega_1(x) \not = 0$,} \ \omega_{x}(y) = \frac{\omega(xy)}{\omega_1(x)}.\]
Then $\omega_1, \omega_x \in \Omega(A)$, and for all $y \in X(A)$, we have $\omega(xy) = \omega_1(x) \omega_{x}(y)$.  We then have 
\[\omega(xy) = \omega_1(x) \omega_{x}(y).\]

{\bf Lemma 9:} {\em Let $A$ be a sequential model, and 
let $e$ be the identity element of the monoid $X(A)$. Then 
$\{e\} \in \M(A)$.} 

{\em Proof:} 
Since we are assuming $\Omega(A)$ is positive, we can find a state
$\omega \in \Omega(A)$ with $\omega(e) > 0$. From the foregoing discussion, we now have 
\begin{equation} 
\omega(x) = \omega(ex) = \omega_1(e)\omega_{e}(x).
\end{equation} 
and 
\begin{equation} 
\omega(x) = \omega(xe) = \omega_1(x)\omega_{x}(e). 
\end{equation}
Summing the first of these over $x \in E$, where $E$ is any test in $\M(A)$,  we have 
$1 = \omega_{1}(e)$. Summing (\theequation) over $E$, we find 
that 
\[1 = \sum_{x \in E} \omega_{1}(x)\omega_{x}(e).\]
Since $\sum_{x \in E} \omega_1(x) = 1$ and $\omega_{x}(e) \in [0,1]$ for every 
$x \in E$, we conclude that $\omega_{x}(e) = 1$ for every $x \in E$, and hence, 
as $E$ was arbitrary, for every $x \in X$. It now follows that $\omega(e) = \omega(ee) = \omega_{1}(e) \omega_{e}(e) = 1$. 
Appealing again to the positivity of $\Omega(A)$, 
we conclude that $\{e\} \in \M(A)$. $\Box$ 

{\bf Corollary 2:} {\em $\omega_1 = \omega$ for every $\omega \in \Omega(A)$. Hence, 
for any $\omega$ and any $x, y \in X(A)$, we have 
\[\omega(xy) = \omega(x)\omega_{x}(y).\]} 

{\bf Proposition 3:} {\em  Sequential models are exactly the 
$(~ \cdot ~)^{c}$ algebras}

{\em Proof:}  Let $A$ be a sequential model with outcome-space $X = X(A)$, and with sequential product $\pi : x,y \mapsto x \ast y$ for $x, y \in X$. Let  $\phi : X^{\ast} \rightarrow X(A)$ be the canonical mapping taking a string of elements of $X(A)$ to their product, defined recursively by 
$\phi(\epsilon) = e$ and $\phi(\a x) = \pi(\phi(\a),x) = \phi(\a)\ast x$ where 
$\a \in X^{\ast}$, $y \in X$, and $\a\b$ represents 
the concatenation of strings $\a$ and $\b$. This map will define a $(~\cdot~)^c$ structure if it is a test-preserving morphism of models. Since 
$\mu$ is the identity on $X(A)$ (understood as a subset 
of $X(A)^{\ast}$), $\phi$ carries $\M(A) \subseteq \M(A^{c})$ to itself. Since $\M(A)$ is inductive in $\Power(X(A))$, its preimage under $\phi$ is an inductive set containing 
$\M(A)$. By Lemma 9, it also contains $\{\epsilon\}$, where $\epsilon$ is the empty string. Thus, $\phi^{-1}(\M(A))$ certainly contains $\M(A^c)$, whence, 
$\phi$ takes tests in $\M(A^{c})$ to tests in $\M(A)$. 
If $\u \perp \v$ in $X^{\ast}$, then  
$\u = \a x \b$ and $\v = \a  y \c$ for some $\a,\b,\c \in X^{\ast}$ 
and some $x, y \in X$ with $x \perp y$, whence, 
$\phi(\u) = \phi(\a)\ast x \ast \phi(\b)$ and $\phi(\v) = \phi(\b) \ast y \ast \phi(\c)$,  are orthogonal in $X$ by condition (3) above.  
This much shows that $\phi$ is a morphism of test spaces. 
{
Finally, if $\omega$ is a state in $\Omega(A)$, then for any $x \in X(A)$, $\omega_{x} \in \Omega(A)$ and, for any 
string $\a \in X^{\ast}$, Corollary 2 gives us 
\[\phi^{\ast}(\omega)(\a,x) = 
\omega(\phi(\a)x) = \omega(\phi(\a))\omega_{\phi(\a)}(x) 
= \phi^{\ast}(\omega)(\a) \omega_{\phi(\a)}(x).\] 
Arguing by induction on the length of strings, we see that $\phi^{\ast}(\omega) = \omega_{f}$ where 
$f(\a,x) = \omega_{\phi(\a)}(x)$: this last is a valid 
transition function with initial and conditional states 
belonging to $\Omega(A)$, and thus, $\omega_{f} \in \Omega(A^{c})$. Thus, $\mu$ is a morphism of models. 

For the converse, suppose $A$ is a $( \cdot )^c$ algebra, 
with unit and multiplication  $\eta : A \rightarrow A^{c}$ 
and $\phi : A^{c} \rightarrow A$. Applying the 
functor $X : \Prob \rightarrow \Set$, we see that $X(A)$ is a monoid with product $x, y \mapsto \phi(xy)$ and identity 
element $\phi(\epsilon)$. Since $\eta : A \rightarrow A^c$ 
is an embedding and $\phi \circ \eta = \id_{A}$,  we see 
that $\phi$ is test preserving, by Lemma 2.  If 
$E \in \M(A)$ and $F : E \rightarrow \M(A)$, then we have 
$\bigcup_{x \in E} \{x\} \times F_{x} \in \M(A^c)$. 
The image of this test under $\phi$ is 
$\bigcup_{x \in E} x F_{x}$, so $\M(A)$ is inductive.  $\Box$

{\bf Non-atomicity of Sequential Models}  If $\M$ is an algebraic test space, 
an event $a \in \Ev(A)$ is {\em atomic} iff the 
corresponding element $[a]$ is an atom in the 
orthoalgebra $\Pi(\M)$. Evidently, $a$ must be 
a singleton, i.e., a single outcome. However, 
not every outcome need be atomic. Indeed, 
$x \in X(A)$ is atomic iff it is not perspective 
to any event $a$ with $|a| > 1$. This condition 
makes sense whether or not $\M$ is algebraic:

{\bf Definition:} An outcome $x \in X(A)$ is 
{\em atomic} iff for all events $a \in \Ev(A)$, 
$x \sim a$ implies $a = \{y\}$ for some outcome $y \in X(A)$. 
The model $A$ is atomic iff every outcome is atomic, 
and {\em totally non-atomic} iff no outcome is atomic. 

As we will now see, under fairly weak conditions on the underlying monoid, sequential models are always totally non-atomic.  We begin with a simple observation: the identity element $e \in X(A)$ is an outcome.

{\bf Proposition 4:} {\em Every sequential model 
 is totally non-atomic.} 

{\em Proof:} Let $A$ be a sequential model. Thus, 
$X(A)$ is a monoid with unit $e$. Let $E \in \M(A)$ 
be any test with $|E| \geq 2$. Since $\{e\}$ is also a  
test, we have $\{x\} = x\{e\} \sim xE$. Since 
left multiplication by $x$ preserves orthogonality, 
if $y, y' \in E$ with $y \not = y'$, $xy \not = xy'$. 
Thus, If $|xE| \geq 2$, and $x$ is not atomic. $\Box$ 

It follows that if a sequential model $A$ 
is algebraic, $\Pi(A)$ is a totally non-atomic orthoalgebra.   In particular, if $A$ is algebraic, so is $A^{c}$, and thus $\Pi(A^{c})$ is a totally non-atomic orthoalgebra. 
Von Neumann models are algebraic, and the logic of a von Neumann model is isomorphic to the projection lattice of the corresponding von Neumann algebra. Since the projection lattices of type-I von Neumann algebras are atomic,  we have 

{\bf Corollary 3:} {\em A von Neumann model associated 
with a type-I von Neumann algebra admits no sequential product. }

A test space $\M$ is {\em semi-classical} iff distinct tests are disoint sets. Such a test space is obviously coherent; as events are perspective iff equal, it also regular and algebraic.  By Proposition 2, $A^c$ inherits these features, so we have 

{\bf Corollary 4:} {\em If $\M$ is semi-classical and all tests in $\M$ have at least two 
outcomes, then $\Pi(\M^{c})$ is a totally non-atomic orthomodular poset.} 
In fact, more is true. In \cite{RJF}, using different methods, it is shown that under the assumptions of Corollary 3, the lattice $C(X(A^c),\perp)$ of orthoclosed subsets of $X(A^{c})$ --- which one can show is isomorphic to $\Pi(A^{c})$ --- is a totally non-atomic complete OML in which every interval is isomorphic to a cardinal power of $L$.

{\em Remark:} It is a perennial concern to try to define some satisfactory ``sequential product" of effects, so that 
the product $a \star b$ of two effects can be read as ``first $a$, and then $b$" \cite{B, LS, GG, dW}. However, 
the standard candidate for quantum effects,  $a \star b = \sqrt{a}b\sqrt{a}$, is not associative,  raising difficulties for the intended interpretation 
 \cite{HHPBS}. The structure of $A^{c}$ may help us better understand why the idea of forming a $\star$ product is inherently problematic. If 
$a, b \in \Ev(A)$ and $c \in \Ev(A^{c})$, then if $a \sim b$, we have 
$ca \sim cb$, but in general $ac \not \sim bc$. Thus, if $A$ is algebraic, 
we have a well-defined {\em action} of $\Ev(A^{c})$ on $\Pi(A)$, namely 
$c[a] = [ca]$, but {\em not} a well-defined product $[a],[c] \mapsto [ac]$. 
Similar remarks apply to the effect algebra $[0,u_A]$ associated with $\E(A)$: the monoid $\Ev(A^{c})$ acts on the the effect algebra, but this 
action generally does not ``linearize" in the first variable to give a sensible sequential product on the latter.   

The moral is the same in both cases: if one wants to 
consider sequential experiments, and particularly if 
one is interested in conditioning, one needs to 
consider the whole  model $A$, and in particular, its operational apparatus $\M(A)$, rather than invariants such as $\Pi(A)$ and $[0,u_A]$ --- a point made 
very clearly by Wright \cite{RW-Spin} almost 50 years ago.

\section{Closing under both $\#$ and $( \cdot )^{c}$.}

Since both compounding and coarse-graining are operationally reasonable ways of synthesizing new experiments from existing ones, one would like to 
be able to close a given model under both constructions 
at once. We have two endo-functors on $\Prob$, 
given on objects by 
by $F(A) = (A^{\#})^{c}$ and $G(A) = (A^{c})^{\#}$. 
As we'll now discuss, $G$ is a monad, supplying the 
desired closure of $A$. Much of what follows 
is routine but somewhat tedious to check. Some details are 
provided in the appendix.

{\bf Lemma 10:} {\em For every model $A$, there is a natural embedding $\ell_{A} : (A^{\#})^{c} \rightarrow 
(A^{c})^{\#}$ given by 
\[(a_1, \cdots, a_n) \mapsto a_1 \times \cdots \times a_n.\]
where the product on the right represents the 
setwise product in $X^{\ast}$ of the sets $a_i \subseteq X^{\ast}$. This defines a natural transformation from 
$F$ to $G$. }

At this point, it will be helpful to recall some 
standard results on the composition of monads. 
If $S$ and $T$ are monads on a category 
$\Cat$, a natural transformation $\ell : ST \nat TS$ is said to be a {\em distributive law} for $T$ over $S$ iff the following 
diagrams commute for all $A$ 
\[
\begin{array}{ccc} 
\begin{tikzcd}
S S T A 
\arrow[d, "\mu^{S}_{TA}" left] 
\arrow[r, "S \ell_{A}"] 
& 
S T S A 
\arrow[r, "\ell_{SA}"]
& T S S A 
\arrow[d, "T \mu^{S}_{A}"]  \\
ST A \arrow[rr, "\ell_{A}"]  & & TSA 
\end{tikzcd}
&  &
\hspace{-.5in} \begin{tikzcd}
& SA  
\arrow[dr, "\eta^{T}_{SA}" above right] 
\arrow[dl, "S \eta^{T}_{A}" above left] 
& \\ 
STA \arrow[rr, "\ell_{A}"] & & TS A 
\end{tikzcd}\\
& \vspace{.1in} & \\
\begin{tikzcd}
S T T A 
\arrow[d, "S \mu^{T}_{A}" left] 
\arrow[r, "\ell_{TA}"] 
& 
T S T (A) \arrow[r, "T \ell_{A}"]
& TTS 
\arrow[d, "\mu^{T}_{SA}"]  \\
ST A 
\arrow[rr, "\ell_{A}"]  & & TSA 
\end{tikzcd} 
 &   & 
\hspace{.2in} \begin{tikzcd}
& TA  
\arrow[dr, "\eta^{S}_{TA}" above right] 
\arrow[dl, "T \eta^{S}_{A}" above left] 
& \\ 
STA \arrow[rr, "\ell_{A}"] & & TS A 
& &  
\end{tikzcd}
\end{array}
\]

\tempout{
\begin{tikzcd}
S T T A \arrow[d, "\mu^{S}_{TA}" left] \arrow[r, ''\ell_{TA}"] 
& 
T S T (A) \arrow[r, "\ell_{SA}"]
& TSS \arrow[d, "T \mu^{S}_{A}"]  \\
ST A \arrow[rr, "\ell_{A}"]  & & TSA 
\end{tikzcd}
& 
\hspace{.2in} 
& 
\begin{tikzcd}
 & SA  \arrow[dr, "\eta^{T}_{SA}" above right] \arrow[dl, "S \eta^{T}_{A}" above left] & \\ 
ST \arrow[rr, "\ell_{A}"] & & TS A 
& &  \\
\vspace{-.2in}
\end{tikzcd}
\end{array}
\]
}
}

For an example, let $S$ and $T$ be the monads on $\Set$ given (on objects) by $SX = X^{\ast}$ and $TX = \Power(X)$. There is a distributive law for $T$ over $S$ specified by letting 
\[\ell_{X} : \Power(X)^{\ast} \rightarrow \Power(X^{\ast})\]
be the mapping sending a string $a_1,...,a_n$ of sets $a_i \subseteq X$ to the set  
$a_1 \times \cdots \times a_n$ of strings $x_1,....,x_n$ with $x_i \in a_i$ for every $i$. 
When estricted to strings of events of a model $A$, the mapping $\ell_{X(A)}$ is precisely the morphism $\ell_{A}$  discussed in Lemma 10, so the following is not entirely surprising:

{\bf Lemma 11:} {\em The morphisms 
$\ell_{A}$ given in Lemma 10 define a distributive law for 
$(~\cdot~)^{\#}$ over $(~\#~)^{c}$. }

This is verified directly, and in gratuitously gory, detail in Appendix D. However, one can also apply the following observation:

Given a distributive law $\ell : ST \rightarrow TS$, 
one can turn $TS$ into a monad by defining 
\[\mu^{TS} = \mu^{T}\mu^{S} T \ell S \ \mbox{and} \  
\eta^{TS} = \eta^{T} \eta^{S}.\]
In terms of components, 
\[
\begin{tikzcd} 
TS TS A \arrow[r, "T \ell_{SA}"] 
& TTSSA \arrow[r, "\mu^{T}_{SA} \mu^{S}_{A}"] 
& TS A
\end{tikzcd}\]
and 
\[\eta^{TS}_{A} : A \stackrel{\eta^{S}_{A}}{\longrightarrow} SA \stackrel{\eta^{T}_{SA}}{\longrightarrow} TSA.\]

{\bf Lemma 12:} {\em If $A$ is a sequential model, then under elementwise multiplication of events, $A^{\#}$ is also a sequential model.} 

{\em Proof:} Since $X(A)$ is a monoid, so is its power set under elementwise multiplication. 
The set $X(A^{\#})$ of non-empty events of $A$ is closed under setwise multiplication, and 
contains the singleton $\{e\}$, so, is again a monoid.  If $a$ is a fixed non-empty event, the mapping $X(A^{\#}) \rightarrow X(A^{\#})$ given by 
$b \mapsto ab$ is orthogonality-preserving: since the monoidal product 
$X(A^c) \rightarrow X(A)$ is a morphism in $\Prob$, it preserves orthogonality of events, 
so $ab \perp ac$.  If $\{a_i\}$ is a partition of a test $E_i$ in $\M(A)$ and, for each $i$, $\{b_{i,k}\}$ is a partition of some $F_i\in \M(A)$, then $\bigcup_{i} a_{i} \{b_{i,k} | k\}$ is a partition of $\bigcup_{x \in E} x F_{x} \in \M(A)$. Thus, 
$\M(A^{\#})$ is inductive.  Since states $A^{\#}$ are essentially the same as those on $A$, this shows that $A^{\#}$ is again a sequential model. $\Box$ 

A morphism $\phi : A \rightarrow B$ of sequential models --- that is, of $(~ \cdot ~)^{c}$-algebras --- is simply a monoid homomorphism $X(A) \rightarrow X(B)$ 
that is also a morphism of models. If this is the case, since $\phi^{\#}(ab) = \phi(a)\phi(b)$, it is also a morphism of sequential models. In other words: $\#$ lifts to a monad on the category  $\Prob^{c}$ of sequential models. One can show that this yields the desired distributive law; see \cite[pp. 313-318]{Barr-Wells} or \cite{Beck}. 

{\bf $G$-Algebras}  
If $S$ and $T$ are monads on a category $\Cat$ and $\ell : ST \rightarrow TS$ is a distributive law, then one can show that 
a $TS$-algebra $(A,\gamma)$ is also both a $T$-algebra
and and $S$-algebra under  under the morphisms 
$\alpha$ and $\beta$ given by 
\[ \begin{tikzcd}
S A \arrow[r, "\eta^{T}_{SA}"]  \arrow[dr,"\alpha"']
& TSA \arrow[d, "\gamma"]
& T (A) \arrow[l, "T \eta^{S}_{A}"'] \arrow[dl, "\beta" ]\\
& A & 
\end{tikzcd} \]
Moreover, $\beta$ distributes over $\alpha$ in the sense that 
\begin{equation} 
\alpha \circ S\beta  = \beta \circ  T\alpha \circ \ell.
\end{equation} 
Conversely, the pair $(A,\alpha)$ and $(A,\gamma)$ determine $\gamma$, and one can identify $TS$-algebra with the structure 
$(A,\alpha,\beta)$ satisfying equation (\theequation).   This is (part of) the contents of \cite[Proposition 2]{Beck}. See also \cite[Proposition 2.1]{Barr} 

As an illustration, consider again the case where $S$  is the free-semigroup monad and $T = \Power$ is the covariant power set monad on $\Set$, 
and let $\ell_{X} : (a_1,...,a_n) \rightarrow a_1 \times \cdots \times a_n$ be the 
distributive law considered earlier.  An $S$-algebra $(X,\alpha)$ is effectively a semigroup 
with $x \ast y := \alpha(x,y)$, and a  
$\Power$-algebra $(X,\beta)$ is essentially a complete lattice with  
$\beta(a) = \bigvee a$. Noting that $\Power \alpha$ is just the usual extension of $\alpha$ to subsets of $X^{\ast}$, 
(\theequation) tells us that 
\[\alpha(\beta(a_1),...,\beta(a_n)) = \beta(\alpha(a_1 \times \cdots \times a_n),\]
or (read from right to left) 
\[\bigvee (a_1 \ast \cdots \ast a_n) = \bigvee(a_1) \ast \cdots \ast \bigvee(a_n). \]
for all $a_i \in \Power(X)$. In other words, $\bigvee$ distributes over setwise products. 
Specializing to the case in which $n = 2$, $a_1 = \{x\}$ and $a_2 = a$, we have 
$x \ast (\bigvee a) = \bigvee (x \ast a)$; similarly if $a_1 = a$ and $a_2 = \{x\}$, 
$(\bigvee a) \ast x  = \bigvee (a \ast x)$, so left and right multiplication by $x \in X$ 
distribute over joins. This makes $X$ a {\em quantale} \cite{Rosenthal}, and, 
conversely, any quantale is an algebras for $\Power \circ S$.

Returning to our monads of interest, $T = ( \cdot )^{\#}$ and $S = (\cdot )^{c}$,  
we see that a $G := TS$-algebra is a model $A$ equipped with both a coherence $\sigma$ 
and a sequential product $\ast$ such that 
\[\sigma(a_1 \ast \cdots \ast a_n) = 
\pi(\sigma(a_1),...,\sigma(a_n)) = \sigma(a_1) \ast \cdots \ast \sigma(a_n)\]
for any events $a_1,...,a_n$ in $\Ev(A)$. In particular, 
given outcomes $x, y \in X(A)$, we have 
\[\sigma(x \ast y) = \sigma(x) \ast \sigma(y).\]

Suppose the $G$-algebra $A$ is  algebraic and strongly unital. 
By Lemma 6, $\Pi(A)$ is an orthomodular poset. 
For any event $a \in \Ev(A)$, we have 
\[p(\sigma(a)) = \oplus\{ [x] | x \in a\}.\]
We also have an action of $\Ev(A)$ on $\Pi(A)$, given by 
\[a \cdot [b] = [a \ast b]\]
for all $a, b \in \Ev(A)$. The term $[a \ast b] \in \Pi(A)$ depends only on the equivalence class $[b]$, but on the specific event $a$. In other words, 
the monoid $\Ev(A)$ acts on $\Pi(A)$ by orthomorphisms.\footnote{This is reminiscent of the dynamic quantum logic of \cite{Baltag-Smets, Kishida-etal}, but I will not pursue this connection here.}

\tempout{
{\bf Example: Classical models as $G$-algebras}   Let $A$ be a classical model based on a measurable space $(S,\Sigma)$. 
This carries both a natural coherence and a natural 
commutative sequential product, given respectively by 
\[\sigma(\{a_i\}) = \bigcup_i a_i\]
and 
\[\pi((a_1,...,a_n)) = \bigcap a_i\]
for all pairwise disjoint families $\{a_i\}$ of non-empty sets 
$a_i \in \Sigma$, and for all strings of sets $a_i \in \Sigma$. Regarding the latter, note that 
$\pi(\emptyset) = S$. Since $\bigcup_i (\bigcap_j a_{ij}) 
= \bigcap_{j}(\bigcup_i a_{i,j})$, the structure 
$(A,\sigma,\pi)$ is a commutative $G$-algebra. (Notice, 
however, that a free $G$-algebra, including 
the free $G$-algebra a classical model $A$, is non-commutative.)
}

It is a well-rehearsed idea that commutativity is a hallmark of classicality. The following result shows that this is 
the case in the present context, at least for algebraic 
$G$-algebras.  If $\M$ is an algebraic test space, one says that one test $E$ {\em refines} another, $F$, iff for every 
$y \in F$ there exists an event $a \subseteq E$ with 
$y \sim a$. It is a standard result that $\Pi(\M)$ is 
a boolean algebra iff every pair of tests have a common refinement. 

{\bf Proposition 5:} {\em Let $A$ be an algebraic, commutative $G$-algebra. Then $\Pi(A)$ is a Boolean algebra} 

{\em Proof:} If $E$ and $F$ are two tests in 
$\M(A)$, then $EF$ is a test as well. We now have 
$a = ae \sim aF$. Since $A$ is commutative, 
$b = be \sim bF = Fb$. Thus, $EF$ is a common 
refinement of $E$ and $F$, and the result follows. $\Box$

{\bf $G$-Algebras and Interference} The interplay between 
the sequential product and a coherence $\sigma$ is generally 
non-trivial.  As mentioned earlier, if $a, b \in \E(A)$ with $a \sim b$, it needn't be the 
case that $ay \sim by$ for a given outcome $y \in X(A)$. 
In particular, if $\omega \in \Omega(\for{AB})$, the probabilities $\omega(\sigma(a)y)$ and $\omega(ay) = \sum_{x \in a} \omega(x,y)$ may differ. 
This is exactly what we see in quantum mechanical spin 
experiments discussed at the end of Section 2. Following 
\cite{RW-Spin}, we therefore make the following 

{\bf Definition:} Let $A$ be a sequential test space, let 
$\sigma$ be a coherence on $A$, 
and let $a \in \Ev(A)$. We say that 
a state $\omega \in \Omega(A)$ {\em exhibits interference} 
between the outcomes $x \in a$ {\em with respect to $\sigma$} \cite{RW-Spin} iff 
for some outcome $y$, $\omega(\sigma(a)y) \not = 
\sum_{x \in a} \omega(xy)$.  In this case, we 
say that $y$ {\em witnesses}, or is a {\em witness for}, this interference.  More generally, 
we say that a model $A$ exhibits interference iff some state of $A$ does so.

{\bf Proposition 6:} {\em $G$-algebras exhibit no interference.}   

{\em Proof:} Since the coherence $\sigma$ of a $G$-algebra is a monoid-homomorphism on 
$\Ev(A)$, and also fixes single outcomes, we have for any event $a$ and outcome $y$ that 
\[ay \sim \sigma(ay) = \sigma(a)\sigma(y) = \sigma(a)y.\]
Hence, for every state $\omega$, $\omega(\sigma(a)y) = \omega(ay) = \sum_{x \in a} \omega(xy)$. $\Box$ 

$G$-algebras, then, while mathematically interesting in other respects, have little directly to tell us about interference: it is the {\em departure} of a structure $(A,\sigma)$ --- $A$ a sequential model and $\sigma$, a coherence on it --- from being a $G$-algebra, that is, the departure of $\sigma$ from being a homomorphism, that enables interference effects.\footnote{This suggests, though at present this is simply speculation, that some sort of cohomology is involved.} It would seem reasonable, therefore, to investigate structures --- let me give them 
the working title of {\em $\#$-modules} --- of the form $(A,\sigma)$, where $A$ is a sequential model and $\sigma$ is a coherence satisfying the weaker equivariance condition $x\sigma(a) = \sigma(xa)$ for events $a \in \Ev(A)$ and outcomes $x \in X(A)$.

\tempout{
Is it automatic that $\sigma$ is a cohesion in this case?
OK, so let $a \oc c \oc b \oc d$ in $\Ev(A)$. 
Perform $a \cup c$; if get $a$, record outcome; if 
get $c$, perform $b \oc d$. So this is 
\[a \cup \cup cb \cup cd\]
This is a test in $\for{AA} \subseteq A$. Now, this is 
the same as $a \cup bc \cup dc = a \cup ((b \cup d)c) = 
a \cup c$. 
\[a(b \cup d) \cup c(b \cup d)
= ab \cup ad \cup cb \cup cd\]

OK: $E = a \cup b$, $F = c \cup d$. Form 
$EF = ac \cup bc \cup ad \cup bd$. This is a 
test in $A$. Now coarse-grain. Have 
$a \sim a(c \cup d)$, $b \sim b(c \cup d)$, 
and similarly (using commutativity!) 
$c \sim c(a \cup b) = ac \cup bc$ and 
$d \sim d(a \cup b) = ad \cup bd$. So 
every pair of tests has a common refinement. 

Next: if $a \perp b$ is $ab = \emptyset$? In commutative 
case? Well, start with $a \co b$, and let 
$E = a \cup b$. Then$a = a1 = aa \cup ab$ and 
$b = bb \cup ba = bb \cup ab$, so 
\[ab = (aa \cup ab)(bb \cup ab) = aabb \cup aaab \cup abbb \cup ab ab.\]
Commutativity makes this $ aabb \cup aaab \cup abbb$ 

Next: $a \oc c \oc b \oc d$: common-refine 
$E = a \cup c$ and $F = b \cup d$    
}

\section{Conclusion and Prospectus}

By constructing a categorical framework for the study of GPTs, we have been able to identify two particularly important operational constructions, coarse-graining and the formation of compound measurements, as monads, and to investigate how these are related, and their respective algebras. In doing so, we've established important links with several themes in the older quantum-logical literature. In particular, we have opened (or rather, re-opens) a path to the study of interference in general probabilistic theories. It would be of interest to compare this approach with that of \cite{Garner-interference},  and with work on higher-order interference in GPTs \cite{Barnum-Lee-Selby, Horvat-Dakic}. 

There are many remaining questions, especially concerning the  category of $G$-algebras and the more general category of ``$\#$-modules".  The following five, presented in no particular order, seem to me to be particularly interesting. 

{\bf Question 1: Projective $G$-algebras.} If $A$ is cohesive and projective,  
then $A$ is algebraic, and hence, so is $G(A)$. What can 
be said about the detailed structure of $G(A)$ and its 
logic, $\Pi(G(A))$? In general, projectivity is lost in passing from a model 
$A$ to either $A^{\#}$ or $A^{c}$. Hence, if $A$ is projective, the free $G$-algebra $GA = (A^{c})^{\#}$ is generally not. A locally finite, projective $G$-algebra would 
essentially be $(\D(L),\Omega)$ for a totally non-atomic OMP $L$.  Several examples of such a thing are known: any non-atomic Boolean algebra; the projection lattice of a type II$_1$ factor, or the homogenous OML  from \cite{RJF} mentioned earlier. But can one find more exotic --- in particular,  non-lattice ordered --- examples?


{\bf Question 2: Tensor Products.}  Do the categories 
$\Prob^{\#}$, $\Prob^{c}$, or $\Prob^{G}$ admit natural 
non-signaling monoidal products? 
If so, 
what can be said about 
compact structures with respect to such products? 
\tempout{
There are several natural products on $\Prob$ that turn the latter into a monoidal category (see \cite{PLock}). 
One is defined on objects by $A, B \mapsto A \times_{NS} B$ where $\M(A \times_{NS} B) = \{E \times F | E \in \M(A), F \in \M(B)\}$, and $\Omega(A \times_{NS} B)$ consists of all states having well-defined conditional (and hence, marginal) states belonging to $\Omega(A)$ and $\Omega(B)$ \cite{LTS}.  It is not hard to see that $\#$ is strong-monoidal over $\times_{NS}$, so $A, B \mapsto A \# B := (A \times_{NS} B)^{\#}$ defines a symmetric monoidal product of $\#$-algebras. In general, this model supports an abundance of 
entangled states, but no entangled outcomes or effects.  Another, simpler, composite is $A, B \mapsto A \amalg B$, where $X(A \amalg B) = X(A) \amalg X(B)$ (the 
disjointified union of $X(A)$ and $X(B)$), $\M(A \amalg B)$ 
is the union of the canonical images of $\M(A)$ and $\M(B)$ 
in $X(A \amalg B)$, and $\Omega(A \amalg B)$ is isomorphic 
in the obvious way to $\Omega(A) \times \Omega(B)$.  
This evidently has no entangled states. \footnote{This is the coproduct in the category of probabilistic models and test-preserving morphisms.}
One finds that $( \cdot )^{c}$ is strong monoidal with respect to $\coprod$, so we can define a tensor product of sequential models by $A, B \mapsto AB := (A \coprod B)^{c}$. \footnote{Note that $\for{AB}, \for{BA} \subseteq AB$, allowing us to define for a probability weight on $AB$,  marginal and conditional states on $A$ and $B$. However, it is not entirely clear how this relates to the usual definition of non-signaling.}
}

{\bf Question 3: von Neumann models.} If $A$ is a von Neumann model, what is the structure of $G(A)$? Is it embeddable in any von Neumann model? 

{\bf Question 4: Interval models.} Let $\E$  be an order-unit space with order-unit $u$, and let $[0,u]$ denote its order-interval.  Let $\M(\E)$ be the test space consisting of all 
finite sets $\{a_i\} \subseteq (0,u]$ with $\sum_i a_i = u$. It is straightforward to show that 
$\bigcup \M(\E) = (0,u]$: if $a \not = \tfrac{1}{2}u$, then $\{a, u - a\} \in \M(\E)$, and 
for $a = \tfrac{1}{2}u$, note that  
$\{\tfrac{1}{2}u, \tfrac{1}{3}u, \tfrac{1}{6}u\}$ also belongs to $\M(E)$. Every 
state on $\E$ --- that is, every positive linear functional $f$ with $f(u) = 1$ --- restricts to 
a probability weight on $\M(\E)$ in an obvious way, so we have a model, $A(\E) = (\M(\E)),\Omega(\E))$ where $\Omega(\E)$ is $\E$'s state space. We then have $\V(A(\E)) \simeq \E^{\ast}$. In finite dimensions, this gives us  $\E \simeq \V(A(\E))^{\ast}$; in particular, in this case all effects on $A(\E)$ are outcomes in $X(A(\E))$.  Let us call a model of the form $A(\E)$ an {\em interval model}. We can now ask: if $A$ is an interval model, when is $A(\E)^{c}$ an interval model? 

{\bf Question 5: Conditional dynamics} Let us define a {\em conditional dynamics} on a 
probabilistic model $A$ to be an affine mapping $\Phi : \Omega(A) \rightarrow \Omega(\for{AA})$ with the property that $\Phi(\alpha)_{1} = \alpha$. 
The idea is that $\Phi(\alpha)_{x}$ is the state of the system conditional on 
the occurence of outcome $x$, if the initial state was $\alpha$. A sequential 
model carries a canonical conditional dyanamics, namely $\Phi(\alpha)(x,y) = \alpha(xy)$. More generally, a conditional dynamics on any model $A$ gives rise to an {\em effect-valued} sequential product of outcomes in $X(A)$, namely, $(x \odot y)(\alpha) := \Phi(\alpha)(xy)$.
Question: When $A = A(\E)$ for a (say, finite-dimensional) order-unit space $\E$, when does the operation $a, b \mapsto a \odot b$ arising from a conditional dynamics $\Phi$ make $[0,u]$ a sequential effect algebra in the sense of \cite{GG}?

As discussed in Appendix A, one can extend the definition of a morphism from a model $A$ to a model $B$ to include certain partial mappings $X(A) \rightarrow X(B)$. Let 
$\PProb$ and $\PProb_1$ denote, respectively, the categories of models and such partial 
morphisms, and of models and test-preserving partial morphisms. One can show that 
both $(~\cdot~)^{\#}$ and $(~\cdot~)^{c}$ extend to endofunctors of 
$\PProb$ and $\PProb_1$, respectively, so they define monads on these larger 
categories. We leave the discussion of partial sequential models, partial 
$\#$-algebras, etc., to future work.

\vspace{.2in}




{\large \bf Appendix A: Supports and Partial Morphisms} 


It is sometimes more convenient to work with what we 
might call {\em partial} morphisms of models. We briefly 
introduce these, and then consider how the results 
obtained above for (total) morphisms extend the the 
more general category of probabilistic models and partial 
morphisms. 
 
Let $\M$ be a test space with outcome-set $X$. A set 
$S \subseteq X$ is a {\em support} \cite{FGR-II, Wilce-handbook} iff 
\[\M_{S} := \{ E \cap S | E \in \M\}\]
is irredundant, in which case we regard it as a 
test space in its own right. If $\alpha$ is a 
probability weight on $\M$, then $\supp(\alpha) = 
\{ x \in X | \alpha(x) > 0\}$, is a support, whence 
the terminology. 

{\bf Lemma A1:} {\em Let $S \subseteq X$. Then 
$S$ is a support if and only if, for all events $a, b \in \Ev(\M)$ 
\[a \sim b \ \mbox{and} \ a \cap S = \emptyset \ \Rightarrow \ b \cap S = \emptyset.\]}

{\em Proof:} Let $a \co c \co b$, and let $E = a \cup c$ and 
$F = c \cup b$. Then 
$a \cap S = \emptyset$ implies $E \cap S \subseteq c \subseteq F$. Since $S$ is a support, $E \cap S = F \cap S$, whence, 
$F \cap S \subseteq c$, and thus $b \cap S = \emptyset$. 

The following will be useful below. 

{\bf Lemma A2:} {\em Let $S$ be a support of 
$\M$, and define $S^{\#} = \{ a \in \Ev(\M) | 
a \cap S \not = \emptyset\}$, and let 
$S^{\ast}$ be the free monoid on $S$, regarded as a subset 
of $X^{\ast}$ Then $S^{\#}$ is a support 
of $\M^{\#}$, and (b) $S^{\ast}$ is a support 
of $\M^{c}$. } 

{\em Proof:} (a) If $A$ and $B$ are pespective evens in 
$\Ev(\M^{\#})$, then $a := \bigcup A$ and $b := \bigcup B$ 
are perspective in $\Ev(\M)$, and 
$A \cap S^{\#} = \emptyset$ iff $a \cap S = \emptyset$. 
The result then follows from Lemma A1. $\Box$ 

(b) We have $S^{\ast} \cap X = S$, so 
$S \cap E \subseteq F \Rightarrow S \cap E = S \cap F$ 
for all $E, F \in \M \subseteq \M^{\ast}$. 
Now suppose $S^{\ast} \cap X^{n}$ is a support for 
$\M^{n}$ (words of length $\leq n$ and tests contained 
in $X^{n}$). If $E', F' \in \M^{n+1}$, then 
$E' = \bigcup_{x \in E} x E_{x}$ and $F' = \bigcup_{y \in F} y F_{y}$ where $E_x, F_y \in \M$ and $E, F \in \M^{n}$. Thus, 
\[S^{\ast} \cap E' = \bigcup_{x \in E \cap S^{\ast}} x (E_{x} \cap S)\]
If this is contained in $F'$, then $E \cap S^{\ast} \subseteq F$ and $E_{x} \cap S \subseteq F_{x}$. Since 
$S$ is a support, $E \cap S^{\ast} = F \cap S^{\ast}$ 
and $E_{x} \cap S = F_{x} \cap S$, and the result 
follows. $\Box$ 

There is an interesting characterization of 
regularity in terms of supports. A proof can be found in 
\cite{Wilce-handbook}.

{\bf Lemma A.3} {\em Let $\M$ be a test space 
with outcome-space $X$. Then the following are equivalent: 
\begin{mlist} 
\item[(a)] $\M$ is regular 
\item[(b)] for every $x \in \bigcup \M$, $X \setminus x^{\perp}$ is a support for $\M$; 
\item[(c)] $X \times X ~\setminus \perp$ is a support 
of $\for{AA}$.
\end{mlist} }


{\bf Lemma A.4} {\em Let $\M$ and $\M'$ be test spaces, with outcome-sets $X$ and $X'$, respectively. If 
$\phi : X \rightarrow X'$ is a morphism and 
$S$ is a support of $\M'$, $\phi^{-1}(S)$ is a 
support of $\M$.} 

{\em Proof:} If $a, b \in \Ev(A)$ and 
$a \sim b$, then $\phi(a) \sim \phi(b)$, so 
$\phi(a) \cap S = \emptyset$ iff $\phi(b) \cap S = \emptyset$. 
Since $\phi(a) \cap S = \emptyset$ iff 
$a \cap \phi^{-1}(S) = \emptyset$, and similarly for $b$, 
the result follows. $\Box$ 

{\bf Lemma A.5} {\em Let $T$ be a support of 
$\M^{\#}$. Then $T = S^{\#}$ for a support $S$ of $\M$.} 

{\em Proof:} Let $\eta : \M \rightarrow \M^{\#}$ be the 
usual embedding, i.e, $\eta(x) = \{x\}$. Let 
$S = \eta^{-1}(T) = \{ x \in X | \{x\} \in T\}$. 
By Lemma A.4, this is a support of $\M$. Let 
$a \in S^{\#}$. Then $\exists x \in a$ with $\{x\} \in T$. 
Hence $\eta(a) = \{\{x\} | x \in a\} \cap T \not = \emptyset$. 
But $\eta(a) \sim \{a\}$ in $\Ev(\M^{\#})$, so 
by Lemma A.1, $\{a\} \cap T \not = \emptyset$, i.e., 
$a \in T$. Conversely, let $a \in T$. Then as 
$\{a\} \sim \eta(a)$, $\eta(a) \cap T \not = \emptyset$, 
whence, there is some $x \in a$ with $\{x\} \in T$, 
i.e., $x \in S$, and so $a \cap S \not = \emptyset$, 
whence, $a \in S$. $\Box$ 

If $S$ is a support of $\M$, then any probabability 
weight $\alpha$ on $\M_S$ extends uniquely to a 
probability weight $\tilde{\alpha}$ on $\M$, 
namely $\tilde{\alpha}(x) = \alpha(x)$ for $x \in S$ 
and $\tilde{\alpha}(x) = 0$ for $x \not \in S$. Henceforth, 
I will simply identify $\alpha$ with $\tilde{\alpha}$ and 
treat $\Pr(\M_{S})$ as the subset of $\Pr(\M)$ consisting 
of probability weights with support in $S$. 

If $\Omega \subseteq \Pr(\M)$, let $\Omega_{S} = \{ \alpha \in \Omega | \supp(\alpha) \subseteq S\}$. 
$A$ is a probabilistic model and $S$ is a 
support for $\M(A)$, we can then define a model $A_{S}$ 
by setting 
$\M(A_S) = \M(A)_{S}$ 
and 
$\Omega(A_{S}) = \Omega(A)_{S}$.
However, in order to maintain our standing assumption that 
all models are positive, it is necessary here to assume 
that for every $x \in X$, there exists a state 
$\alpha \in \Omega_{S}$ with $\alpha(x) > 0$. In other words, 
$S = \bigcup_{\alpha \in \Omega_{S}} \supp(\alpha)$. 
A support of this kind is said to be {\em stochastic} \cite{FGR-II}.\footnote{Remarkably, by a result of Cohen and Svetlichny \cite{Cohen-Svetlichny}  there exist non-stochastic supports even in the otherwise very well-behaved test space ${\mathcal F}(\H)$ of frames of a Hilbert space.}

{\bf Definition:} Let $A$ and $B$ be probabilistic models. A {\em partial morphism} from 
$A$ to $B$ is a partial mapping from 
$X(A)$ to $X(B)$, with domain $S$, such that 
\begin{mlist} 
\item[(a)] $x \perp y$ in $S$ implies $\phi(x) \perp \phi(y)$; 
\item[(b)] $\phi(E \cap S) \in \Ev(B)$ for every $E \in \M(A)$; 
\item[(c)] $\phi(E \cap S) \sim \phi(F \cap S)$ for all 
$E, F \in \M(A)$; 
\item[(d)] For every $\beta \in \Omega(B)$, $\phi^{\ast}(\beta) \in \V(A_S)$ --- that is, $\phi^{\ast}(\beta)$ is a 
multiple of a state in $\Omega(A_{S})$. 
\end{mlist}  

The following is sometimes useful: 

{\bf Lemma A.6} {\em If $\phi : A \rightarrow B$ is a partial 
mapping $X(A) \rightarrow X(B)$ with domain $S \subseteq X(A)$, satisfying (b) and (c). Then $S$ is a 
support of $\M$. }

{\em Proof:} It follows from (b) and (c) that if $a \sim b$ in $\Ev(A)$, then 
$\phi(a) \sim \phi(b)$ in $\Ev(B)$. If $a \cap S \not = \emptyset$, then $\phi(a) \not = \emptyset$; since $\phi(b) \sim \phi(a)$, $\phi(b) \not = \emptyset$ either, and 
it follows from Lemma A.1 that $S$ is a support. $\Box$ 

If $\phi : A \rightarrow B$ and $\psi : B \rightarrow C$ 
are partial morphisms with domains $S \subseteq X(A)$ and 
$T \subseteq X(B)$, then $\psi \circ \phi$ is a 
partial morphism $X(A) \rightarrow X(B)$ with domain 
$R := S \cap \phi^{-1}(T)$. 
We need to check that $R$ is a support.
If $E \cap R  \subseteq F$, then 
\[S \cap (E \cap \phi^{-1}(T) ) \subseteq F\]
whence, 
\[\phi(\phi^{-1}(T) \cap E) = T \cap \phi(E) \subseteq \phi(F).\]
{
If $\phi(E) \sim \phi(F)$, then $T \cap \phi(E) = T \cap \phi(F)$. So partial morphisms compose. This gives 
us a category, $\PProb$, of probabilistic models and partial 
morphisms. We write $\PProb_1$ for the sub-category of probabilistic models and test-preserving partial morphisms.}

{\bf Examples} Suppose $A$ and $B$ are full Kolmogorovian models, associated with measureable spaces $(M,\sigma)$ and
$(M',\Sigma')$, respectively. Let $f : M \rightarrow M'$ be measurable, and set 
\[S = \{ a \in \Sigma' \setminus \{\emptyset\} | f^{-1}(a) \neq \emptyset\} = \{ a \in \Sigma' | a \cap \ran(f) \not = \emptyset\}.\]
Then we have a mapping $\phi : S \rightarrow \Sigma \setminus \{\emptyset\} = X(M,\Sigma)$, namely $\phi = f^{-1}|_{S}$, and 
it is straightforward to verify conditions (a)-(d) in the 
definition of a morphism $B \rightarrow A$. 
This gives us a contravariant functor from the category of measurable spaces and mappings into $\PProb$. 
 
Similarly, if $\A$ and $\B$ are von Neumann algebras and 
$\phi : \A \rightarrow \B$ is any normal $\ast$-homomorphism, 
let $S = \P(\A) \setminus \ker(\phi)$. It is 
easy to check thtat $S$ is a support and $\phi|_{S}$ is a morphism $\M(\A)_{S} \rightarrow \M(\B)$. 
So we have a covariant functor from the category of 
von Neumann algebras and normal $\ast$-homomorphisms into $\PProb$. 

It is not hard to see that $(~\cdot~)^{\#}$ extends to an endofunctor on $\PProb$. 
As $\Prob$ is a subcategory of $\PProb$ having the same objects, every monad $T$ on $\Prob$ that extends to an endofunctor on $\PProb$ is also a monad on the latter, and every $T$-algebra in $\Prob$ is still a $T$-algebra in $\PProb$.  However, it is possible that there will exist new {\em partial} $T$-algebras $(A,\phi)$ in which the product morphism $\phi : TA \rightarrow A$ is partial. For $T = \#$, this does not happen:

{\bf Lemma A.7:} {\em A partial $\#$-algebra in $\PProb$ is 
total, i.e., a $\#$-algebra in $\Prob$.} 

{\em Proof:} Let $\alpha :  \rightarrow X(A)$ be a 
partial morphism from $A^{\#}$ to $A$. By Lemma A.5, 
$\dom(\alpha) = S^{\#}$ for some support $S$ of $A$. 
In particular, $\alpha$ is defined only for 
events $a$ with $a \cap S \not = \emptyset$. But now 
we have $\alpha \circ \eta_{A} = \id_{A}$, i.e., 
for every $x \in X$, $\alpha(\{x\}) = \{x\}$. It 
follows that $S = X$ and hence, $\dom(\alpha) = X^{\#} = 
\Ev(A) \setminus \{\emptyset\} = X(A^{\#})$. $\Box$ 

There is no corresponding result for the compounding monad. Even for classical models $(\{E\},\Omega)$, any partial monoid on $E$ is a $( \cdot )^{c}$-algebra.

{\bf Event-valued morphisms} 
In the work of Foulis, Randall and some of their students, \cite{FR-MMQM, FPR, PLock} the preferred morphisms $\M(A) \rightarrow \M(B)$ were so-called {\em interpretations}: event-valued mappings $\phi : X(A) \rightarrow \Ev(B)$ 
such that if $x \perp y$ in $X(A)$, $\phi(x) \perp \phi(y)$ 
in $\Ev(B)$, and $\bigcup_{x \in E} \phi(x) \in \M(B)$ 
for every $E \in \M(A)$. If $\beta$ is a probability 
weight on $\M(B)$, $\phi^{\ast}(\beta) = \beta \circ \phi$ 
is then a probability weight on $\M(A)$, so we can extend 
the definition to say that an interpretation 
$\phi : A \rightarrow B$ is an interpretation in the 
above sense from $\M(A)$ to $\M(B)$, with the added 
feature that $\phi^{\ast}$ takes $\Omega(B)$ into 
$\Omega(A)$. 
When $\phi(x) \not = \emptyset$ 
for all $x \in X(A)$, the interpretation is said to be 
{\em positive}. A positive interpretation is a morphism in our sense from $A$ to $B^{\#}$, so the category of test spaces and positive interpretations is the Kleisli category of $\Mod$ under $\#$.  More generally, interpretations 
are test-preserving partial morphisms $A \rightarrow B^{\#}$.  Our partial morphisms $A \rightarrow B^{\#}$ are what Foulis and Randall termed {\em conditionings}. Partial 
test-preserving morphisms $A \rightarrow B^{\#}$ are interpretations. \\

{\bf Pointed Partiality}  By a {\em pointed} model, I mean a structure 
$(A,\ast_{A})$ where $A$ is a probabilistic model, 
and $\ast \in X(A)$ is a distinguished ``null" or ``failure" outcome, such that there exists a unique state $\delta \in \Omega(A)$ with $\delta(\ast) = 1$. Given any model $A$ and any $\ast_{A} \not \in X(A)$, we can form 
a pointed model $A^{+}$ by setting 
\[\M(A^{+}) = \{ E \cup \{\ast_{A}\} | E \in \M(A)\}\]
and 
\[\Omega(A) = \{ t\alpha + (1 - t)\delta | \alpha \in \Omega(A)\}\]
where $\delta : X(A) \cup \{\ast\} \rightarrow [0,1]$ is 
identically $0$ on $X(A)$ and $1$ at $\ast_{A}$. 
Geometrically, $\Omega(A^+)$ is a truncated cone 
with base $\Omega(A)$ and apex at $\delta$. \footnote{One of the great advantages of footnotes is that they provide a safe outlet for pedantic impulses. 
As an illustration of this, let me point out that this construction defines $A_+$ only up to the choice of an object $\ast_{A}$ for every object $A \in \Prob$. One could do this systematically by defining, say, $\ast_{A} = A = (\M,\Omega)$, and stipulating that no model is permitted to include its own ``name" among its outcomes. }

{\bf Lemma A.8:} {\em $A \mapsto A^{+}$ is the object part of a monad on $\Prob$.  The category $\PProb$ is the Kleisli category $\Prob_{+}$, and the category of probabilistic models and general set-valued morphisms is $\Prob_{+\#}$.   }


{\em Proof:} Immediate, since $+$ is a monad on $\Set$. $\Box$. 

A morphism of pointed models from, say, $(A,\ast_A)$ to $(B,\ast_B)$, is 
a morphism of models $\phi : A \rightarrow B$ wi
$\phi(\ast_{A}) = \ast_{B}$. Given such a morphism, 
let $S = X(A) \setminus \phi^{-1}(\ast_{B})$, 
and define 
\[\phi_{S}(x) = \left \{ \begin{array}{cc} \phi(x) & x \in S\\
\ast_B & x \not \in S
\end{array} \right.\]
It is easy to check that $S$ is a support, and $\phi_{S}$, a 
partial morphism. 

Conversely, if $\phi : A \rightarrow B$ is a partial morphism 
with domain $S$, we can extend $\phi$ to a total morphism 
$\phi^{+} : A  \rightarrow B^{+}$ by setting 
$\phi(x) = \ast$ for all $x \not \in S$. In particular, 
$\phi(\ast) = \ast$.  Conversely, if $\phi : A \rightarrow B^{+}$ is a total morphism, let $S = X(A) \setminus \phi^{-1}(\ast)$, and let $\phi_{S} = \phi|_{S}$; then 
$\phi_{S}$ is a partial morphism $A \rightarrow B$ 
with $\phi_{S}^{+} = \phi$.  If $\phi : S  \rightarrow X(B)$ 
is a partial morphism with domain $S$, then 
$\phi^{+ - 1}(\ast) = S$ and $\phi^{+}_{S} = \phi$. Thus, 
there is a bijection between partial morphisms 
from $A$ to $B$, and total morphisms $A \rightarrow B^{+}$. 

It is easy to check that $(\phi \circ \psi)^{+} = 
\phi^{+} \circ \psi^{+}$ for partial morphisms 
$\psi : A \rightarrow B$ and $\psi : B \rightarrow C$; hence, 
we have a functor $\PProb \rightarrow \Prob^{+}$, which 
by the foregoing remarks is an isomorphism of categories. \\


{\large \bf Appendix B:  More on Forward Products} 

Let $\phi : A \rightarrow A'$ 
and, for every $x \in X(A)$, $\psi_{x} : B \rightarrow B'$ 
be morphisms. Define a mapping 
\[(\phi;\psi) : X(A) \times X(B) \rightarrow X(A') \times X(B')\]
by 
\[(\phi;\psi)(x,y) = (\phi(x), \psi_{x}(y)).\]
One might hope that this would again be a morphism, but 
this is not necessarily the case.  For a simple example, let $\M(A) = \{E, F\}$ where $E \cap F = \emptyset$, and let $\M(B) = \{E', F'\}$ where $E' = E \cup \{x\}$ and $F' = F \cup \{x\}$ for some $x \not \in E \cup F$. Let $\phi : X(A) = E \cup F \rightarrow X(B) = E' \cup F'$ be the inclusion map. Since $E \sim F$ in $\Ev(B)$, this is a morphism. Now consider 
$E \times F$ and $F\times F \in \M(\for{AB})$: any complement in $\Ev(\for{AB})$ 
for the former contains $E \times \{w\}$, and any complement for the latter contains 
$F \times \{w\}$. No test in $\M(\for{AB})$ contains both of these, so $E \times E$ and $F \times F$ have no complement in common, and are thus not perspective. 

\tempout{
The following shows that $(\phi;\psi)$ need not even be a weak morphism. Maybe 
better ex. than above?

{\bf Example:}  Let $\M = \{E,F\}$ with $E = \{a,x\}$ and $F = \{x,b\}$, and 
let $\M' = \{E',F'\}$ with $E' = \{a,x,z\}$ and $F' = \{z,x,b\}$. Note that 
$a \sim b$ in $\M$ and in $\M'$. Let $\phi$ and $\psi_{x}$ both be the inclusion 
mapping.   Note that $aa \sim ab$ in $\for{\M\M}$ (both are complementary 
to $ax, ca, cx$: the former gives $EE$, the latter, $aF \cup cE$). 

Let $\alpha(a) = \alpha(b) = 1$ and $\alpha(c) = \alpha(z) = 0$: this 
defines a state on both $\M$ and $\M'$.  Let $\beta_{a}(z) = 1$ (so 
$\beta_{a}(x, a, b) = 0$) and $\beta_{b}(x) = 1$ (so $\beta(z,a,b) = 0$). 
Let $\alpha(a) = \alpha(b) = 1$. Consider the events 
$A = \{aa, ax\}$ and $B = \{bb, bx\}$ in $\for{\M\M}$. Note these 
are perspective, with complment $xE$; but their images are 
not perspective: the complement for $A$ in $\for{\M\M}$ is 
$az \cup \{x,z\}E'$ and that of $B$ is $\{bz\} \cup \{x,z\}E'$. 
Now note that $(\alpha;\beta)(A) = \alpha(a)\beta_{a}(a) + \alpha(a)\beta_{a}(x) = 0$, 
while $(\alpha;\beta)(B) = \alpha(b)\beta_{b}(b) + \alpha(b)\beta_{b}(x) = 1$. 
It follows that $(\alpha;\beta)$ does not pull back to a probability weight 
on $\for{\M\M}$. 
}

Our goal is to characterize those pairs $(\phi,\psi)$ for which $(\phi;\psi)$ {\em is} a morphism. Evidently, the problem in the preceding example is that $\phi$ is not test-preserving. This is part of the general story, but the whole of it.

{\bf Definition:} The {\em core} of a test space $\M$ is 
the event $\cor(A) := \bigcap \M$. 

Evidently, $\M$ is classical iff its core is all of $X = \bigcup \M$. For virtually all non-classical test spaces that arise in practice, the core will be empty.  

{\bf Proposition B1:} {\em 
With notation as above, 
$(\phi;\psi)$ is a morphism iff $\psi_{x}$ is test-preserving for all outcomes $x \in X(A)$ such that $\phi(x) \not \in \cor(\phi(\M(A)))$.} 

Note that if $A$ is classical $\cor(\phi(\M)) = \phi(\M)$ and the theorem places 
no constraint on $(\phi;\psi)$. The proof depends on an independently useful characterization of perspectivity in forward products. 

{\bf Lemma B1:} {\em Let $a = \bigcup_{x \in a_o} x a_x$ and $b = \bigcup_{y \in b_o} y b_y$ 
be events in $\for{AB}$. Then $a \co b$ iff (i) 
$a_o \cup b_o \in \Ev(A)$; (ii) $a_{x} \co b_{x}$ in $\Ev(B)$ for all $x \in a \cap b$, 
and (iii) $a_{x}, b_{y} \in \M(B)$ for all $x \in a_o \setminus b_o$ and $y \in b_o \setminus a_o$. }

{\em Proof:} If $a \co b$, then $a \cap b = \emptyset$, whence, 
if $z \in a_o \cap b_o$, $a_z \cap b_z = \emptyset$, and $a \cup b \in \M(\for{AB})$. Thus, letting $a_x = \emptyset$ 
for $x \in b_o \setminus a_o$ and similary for $b_x $, we have 
\[a \cup b = \bigcup_{x \in a_o \cup b_o} x (a_x \cup b_x)  \in \M\]
whence, $a_o \cup b_o \in \M(A)$, and $a_x \cup b_x \in \M(B)$ for all 
$x \in a_o \cup b_o$. If $x \in a_o \setminus b_o$, then $b_x = \emptyset$, 
so $a_x \in \M(B)$, and similarly $b_x \in \M(B)$ if $x \in b_o \setminus a_o$. 
The converse is clear. $\Box$ 

{\bf Lemma B2:} {\em  Let $a = \bigcup_{x \in a_o} x a_{x}$ and $b = \bigcup_{y \in b_o} y b_{y}$ be events in $\Ev(\for{AB})$. 
Then the following are equivalent: 
\begin{itemize} 
\item[(a)] $a \sim b$ in $\Ev(\for{AB})$ 
\item[(b)] 
\begin{itemize} 
\item[(i)] $a_o \sim b_o$ for every $z \in a_o \cap b_o$, 
\item[(ii)] $a_{z} \sim b_{z}$, and 
\item[(iii)] for any $x \in a_o \setminus b_o$ and $y \in b_o \setminus a_o$, 
$a_x, b_y \in \M(B)$. 
\end{itemize} 
\end{itemize} 
}

{\em Proof:} Suppose (b) holds. Let $c'$ be an axis for $a_o$ and $b_o$, 
and for all $z \in c'$, let $c_{z} \in \M$.  Let $c_1 = \{ z \in a_o \cup  b_o | a_{z}, b_{z} \not \in \M(A)\} \subseteq a_o \cap b_o$ (noting that because $a_{z} \sim b_{z}$, if $a_{z} \not \in \M$, then $b_{z} \not \in \M$ and vice versa). For every $z \in c_1$, let $c_{z}$ be an axis for $a_{z} \sim b_{z}$.  Let $c = \bigcup_{z \in c_1} z c_{z} \cup 
\bigcup_{z \in c_1} z c_{z}$. Then $c_o = c_1 \cup c'$, so 
$c_o \cup a_o = c' \cup a_o \in \M(A)$, and for every $z \in c_o \cap a_o = c_1$, 
$a_z \co c_z$, while for every $x \in a_o \setminus c_o = a_o \setminus c_1$, 
$a_{x} \in \M$, and for every $z \in c_o \setminus a_o = c'$, $c_z \in \M$. 
Thus, by Lemma B1, $a \co c$.  Exactly parallel arguments show that $b \co c$, 
so $a \sim b$. 

For the converse, suppose (a) holds, and let $c = \bigcup_{z \in c_o} z c_{z}$ be a 
complement for both $a$ and $b$. By Lemma 1, we have $a_o \cap c_o = b_o \cap c_o$ 
and for all $x \in a_o \setminus c_o$ and all $y \in b_o \setminus c_o$, 
$a_x, b_y \in \M$. Now define $c_1$ as above, and note that $c_1 \subseteq a_o \cap b_o$, 
and that for all $z \in c_1$, $a_{z} \co c_z \co b_z$, so $a_z \sim b_z$. Let 
$c' = c_o \setminus a_o = c_o \setminus b_o$: then $c' \cap a_o = a' \cap b_o = \emptyset$, and $a_o \cup c' = a_o \cup c_o \in \M$ and $b_o \cup c' = b_o \cup c_o \in \M$, so $a_o \sim b_o$ with axis $c'$. For all $x \in a_o \setminus b_o \subseteq a_o \setminus c_1$, $a_x \in \M(A)$; similarly, $b_y \in \M$ for all $y \in b_o \setminus a_o$. 
For $z \in a_o \cap b_o$, either $a_z, b_z \in \M$, in which case $a_z \sim b_z$ automatically, or $z \in c_1$ and $a_z \sim b_z$ by hypothesis. Thus, all of the conditions 
in (b) are satisfied. $\Box$

{\bf Corollary B1:} {\em If $a, b \in \Ev(\for{AB})$ and, 
for some $x \in a \setminus b$, $a_{x} \not \in \M(B)$, 
then $a \not \sim b$.}

{\bf \em Proof of Proposition B1:} To see that the condition 
is necessary, let $x \in X(A)$ with
$\phi(x)$ not belonging to $\cor(\phi(\M))$, and 
suppose $\psi_{x}$ is not test-preserving. We can 
find some test $F \in \M(A)$ with 
$\phi(x) \not \in \phi(F)$, and we can find some 
test $E$ with $\psi_{x}(E)$ not a test in $\M'$. Now 
consider $a = EE$ and $b = FF$: these are perspective 
in $\for{AB}$, but as $\phi(x) \in \phi(E) \setminus \phi(F)$ and $\psi_{x}(E) \not \in \M(B)$, the corollary to 
Lemma B2 tells us that $(\phi;\psi)(EE) = \bigcup_{z \in E} \phi(z)\psi_{z}(E)$ 
is not perspective to $(\phi;\psi)(FE) = \bigcup_{y \in F} 
\phi(y)\psi_{y}(F)$. 

For the converse, suppose that $\psi_{x}$ is test-preserving 
for all $x$ with $\phi(x) \not \in \cor(\phi(\M))$. 
Evidently, $(\phi;\psi)$ 
is locally injective and event-preserving, so we need only check that it preservs perspectivity. Let $a = \bigcup_{x \in a_o} x a_{x}$ and $b = \bigcup_{y \in b_o} y b_{y}$ be perspective in 
$\Ev(\for{AB})$. we have 
$(\phi;\psi)(a) = \bigcup_{x \in \phi(a_o)} \phi(x) \psi_{x}(a_{x})$ and similarly for $b$. If $\phi(x) \in \phi(a_o) \setminus \phi(b_o)$, then (i) $x \in a_o \setminus b_o$, and (ii) $\phi(x) \not \in \cor(\phi(\M))$. By Lemma 2, 
$a_{x} \in \M(A)$, and, as $\psi_{x}$ is test-preserving, 
$\psi_{x}(a_{x}) \in \M(A')$. Similarly, if 
$y \in F$ with $\phi(y) \in \phi(a_o) \setminus \phi(b_o)$, $\psi_{x}(b_x) \in \M(A')$. It remains to show that 
$\psi_{x}(a_x) \sim \psi_{x}(b_x)$ for every $\phi(x) \in a_o \cap b_o$. But since $\phi$ is locally injective, $x \in a_o \cap b_o$, 
so $a_{x} \sim b_x$ by Lemma B2, and the result follows 
since $\psi_{x}$ is a morphim. Thus, again by Lemma B2, 
$(\phi;\psi)(a) \sim (\phi;\psi)(b)$. $\Box$ 
\\

{\large \bf Appendix C:  Further Properties of Compoundings} 

In this appendix, we supply the proof that $A^{c}$ inherits various conditions on $A$, 
viz., unitality, strong unitality (modulo a mild further constraint), regularity, coherence, and algebraicity. 

We begin with the proof of Lemma 8, which for convenience we repeat: 

{\bf Lemma 8 (bis):} {\em Let $A$ and $B$ both be unital,strongly unital, coherent, regular,  or algebraic. Then $\for{AB}$ has the same property.} 

{\em Proof:} For the proof that $\for{AB}$ is algebraic if $A$ and $B$ are, see \cite{PLock}. If $A$ and $B$ are unital, then for any $x \in X(A)$, $y \in X(B)$, 
we can find states $\alpha \in \Omega(A)$ and $\beta \in \Omega(B)$ with 
$\alpha(x) = \beta(y) = 1$. Then the state $\alpha \otimes \beta \in \Omega(\for{AB})$ 
has value $1$ at $(x,y)$.  

If $A$ and $B$ are strongly unital and $xy \in X \times Y$, let $\alpha(x) = 1$ and $\alpha(u) > 0$ for all $u \not \perp x$. Let $\beta_{x}(y) = 1$ and $\beta_{x}(v) > 0$ for all $v \not \perp y$. For all $u \not = x$, let $\beta_{u} > 0$ on $Y$. Then $(\alpha;\beta)(x,y) = 1$ and $(\alpha;\beta)(u,v) > 0$ if $(u,v) \not \perp (x,y)$.  

If $A$ and $B$ are regular, let $S = \bigcup_{u \in X} u T_{u}$ with $S = X \setminus x^{\perp}$ and $T_{x} = Y \setminus y^{\perp}$ and $T_{u} = Y$ for all $u \not = x$. Again, this is $(X \times Y) \setminus (xy)^{\perp}$, so, 
$\for{AB}$ is regular. 

Let $A$ and $B$ be coherent. Let $C = (\bigcup_{x \in a} x b_x)$ and suppose $(u,v) \in C^{\perp}$, then for all $x \in a$, either $u \perp x$ or $u = x$ and $y \in b_{x}^{\perp}$. 
In the first place, $u \perp a$; in the second, $u \in a$ and $v \perp b_{u}$.  In the first case, $a' = a \cup \{u\}$ is an event, and so therefore is 
$\bigcup_{x \in a'} x' b_{x'}$ where $b_{u}$ is any event containing $v$, so $(u,v) \perp C$. In the second case,  enlarge $b_{u}$ to contain $v$. $\Box$ 

We can define the free monoid on $X$ concretely as follows. For $n \in \N$ with $n > 0$, l
et $X^{n}$ be the set of functions $\x : \{1,...,n\} \rightarrow X$, interpreting $\x \in X^{n}$ as the string $\x  = (x_1,....,x_n)$ with $x_i = x(i)$. Set $X^{\ast}  = \bigcup_{n \in \N} X^{n}$ where $X^0 = \{\e\}$, with $\e \not \in X^{n}$ for any $n$, and with concatenation defined by 
\[\u,\v = (u_1,...,u_n), (v_1,...,v_k) \mapsto \u\v = (u_1,...,v_n, u_1,....,v_k)\]
for $\u \in X^n, \v \in X^{k}$ with $n, k > 0$, and  with $\e \u = \u = \u\e$ for $\u \in X^n$. Let $X_{n} = \bigcup_{k=0}^{n} X^{k}$, the set of strings of length $\leq k$. 

Now define, for every $n$, a model $A^n$ as follows. First, 
\[\M(A^{n}) \ := \ \{ E \in \M(A^c) ~|~  E \subseteq X_n\}.\] 
This is the set of tests having at most $n$ stages. Note that $X(A^{n}) = X_{n}$, the set of strings of length $\leq n$.  Also note that $\M(A^{n}) \subseteq \M(A^{n+1})$ for all $n$ and that $\M(A^c) = \bigcup_{n} \M(A^{n})$.   
Secondly, if $\omega \in \Omega(A^c)$, then $\omega$ restricts to a probability 
weight on $\M(A^{n})$. Define $\Omega(A^{n})$ to be the set of restrictions to $X_n$ of states $\omega \in \Omega(A^{c})$.

Let $\phi_n : X \times X_n \rightarrow X_{n+1}$ be given by $\phi(x,\b) = x\b$ (concatenate). This is injective, and its range is $X_{n+1} \setminus \{\e\}$.  It is easy to check that this defines an embedding 
\[\M(\for{A A^{n}}) \rightarrow \M(A^{n+1}).\]

If $\M$ is any test space let $\M'$ be defined as in 
Appendix A. That is, $\M' = \M \cup \{\{e\}\}$ where 
$e \not \in X = \bigcup \M$. Note that probability 
weights on $\M$ extend uniquely to probability weights 
on $\M'$. We define $\M(A') = \M(A)'$ and 
$\Omega(A') = \Omega(A)$, with elements of the latter 
understood as extended to $\M(A)'$.   

{\bf Lemma C0:} {\em 
In terms of the notation 
introduced earlier, 
\[A^{1} \simeq A^{0} \cup \for{A^0 A} = \for{\{\e\} A} \simeq A'.\]
and 
\[A^{n+1} \simeq \for{A' A^{n}}.\]
}

{\bf Lemma C1:} {\em For any model $A$, if $A$ is unital, 
strongly unital, regular, coherent, or algebraic, then $A'$ has the same 
feature. }

{\em Proof:} The claim regarding unitality is obvious. To see that 
$A'$ is strongly unital if $A$ is, simply note that $e^{\perp} = \emptyset$. 

If $A$ is regular, if $a, b \in \Ev(A)$ are distinct, non-empty events with $a \sim b$, then either $a, b \in \Ev(A)$, or one is a test in $\M(A)$ and the other is $\{e\}$. In the former case, $a^{\perp} = b^{\perp}$ by regularity of $\M(A)$; in the latter $a^{\perp} = b^{\perp} = \emptyset$. 
 
Similarly, if $a \subseteq b^{\perp}$, then if $b^{\perp} = \emptyset$, $a = \emptyset$ and thus $a \perp b$ vaccuously, and if $b^{\perp}$ is non-empty, then $b \not = \{e\}$, so $a, b \in \Ev(A)$, and $a \perp b$ by coherence of $A$. 

If $A$ is algebraic and $a \sim b$, $b \perp c$, then 
if $c = \emptyset$, $a \perp c$ trivially, and if not, 
then either $b = \emptyset$, whence $a = \emptyset$ and 
again $a \perp c$, or $b, c \in \Ev(\M) \setminus \M$, 
whence, $a \in \Ev(\M)$, and $a \perp c$ by algebraicity 
of $A$. $\Box$ 

{\bf Lemma C2:}{\em  Let $A$ be unital, regular, coherent, or algebraic. Then so is $A_n$ for every $n \in \N$. If $A$ is strongly unital and has a strictly 
positive state (e.g., if it is SU and $\M(A)$ has a 
finite test), then $A_n$ is strongly unital for every $n$. } 

{\em Proof:} By induction on $n$. The case $n = 1$ is dealt with by Lemma B1, since $\for{A}^{1} \simeq  A'$. Now suppose the claim holds for $n$: then it holds for $\for{A\for{A}^{n}}$ by Lemma 8. Now, $\phi : \for{A\for{A}^{n}} \rightarrow \for{A}^{n+1}$ is an isomorphism onto $A_{n+1} \setminus \{\{e\}\}$, so the latter is unital, SU, regular, coherent, or algebraic if $A$ is. But $A_{n+1} \simeq \phi(A_n)'$, so by Lemma B1, and we see that $A_{n+1}$ itself is unital, strongly unital etc.  $\Box$ 

We shall say that a model $A$ is a {\em submodel} 
of a model $B$, writing $A \leq B$, iff $X(A) \subseteq X(B)$, 
$\M(A) = \{ E \in \M(B) | E \subseteq X(A)\}$ and 
for all $x, y \in X(A)$, $x \perp_A y$ iff $x \perp_{B} y$, 
and $\Omega(A)$ consists of all restrictions to $X(A)$ of states on $\Omega(B)$.  (This tells us that the inclusion mapping $X(A) \rightarrow X(B)$ is an embedding, 
but the converse is false. So this is a rather strong notion of sub-model.) 

Let $(A_n)$ be a sequence of models with $A_{n} \leq A_{n+1}$. Write $X_n$ for $X(A_n)$, and similarly for $\M_n$ and $\Omega_n$.  We define an inductive limit  model 
$A = \bigcup_n A_n$  by setting $X(A) = \bigcup_n X_n$, $\M(A) = \bigcup_n \M_n$. 
To define the states, note that if $(\alpha_n)$ is a sequence of states $\alpha_n \in \Omega_n$ with $\alpha_n = \alpha_{n+1}|_{X_n}$ (and, by submodel-hood, such a 
sequence always exists), then $\alpha(x) := \alpha_n(x)$ where $x \in X_n$ defines a probability weight on $\M(A)$. Let $\Omega(A)$ be the collection of all probability weights of this form.

{\bf Lemma C3:} {\em Let $A_{n} \leq A_{n+1}$ for all $n$ with $A = \bigcup_n A_n$.
\begin{mlist} 
\item[(a)] If, for every $n \in \N$, $A_n$ is unital, strongly unital, regular,  or algebraic, 
then  so is $A$. 
\item[(b)] If $A_n$ is coherent and locally finite for every $n \in \N$, then 
$A$ is also locally finite and coherent.
\end{mlist} 
}

{\em Proof:} (a) This is clear for unitality and strong unitality. If $a \sim b$ in $\Ev(A)$ and $x \in X(A)$, then 
for some $n$, $a \sim b \in \Ev(A_n)$ and $x \in X(A_n)$. 
If $x \perp a$ in $A$, then $x \perp a$ in $A_n$, 
so $x \perp b$ by regularity. If $a, b, c, d$ are events in $\Ev(A)$ with $a \co c \co b \co d$, then 
for some $n$, $a \co c \co b \in \Ev(A_n)$; thus, if $a \sim b \co d$ in $\Ev(A)$ implies 
$a \sim b \co d$ in some $A_n$. Since $A_n$ is algebraic, $a \co d$ in $\Ev(A_n)$, whence, $a \co d$ also in $\Ev(A)$, and $\M(A)$ is algebraic. 

(b) If every test $E \in \M(A_n)$ is finite for every $n \in \N$, then every test, and 
hence, every event, of $A$ is finite. It  $b \subseteq a^{\perp}$ in $\Ev(A)$, then, because $a$ is finite,   $b \subseteq a^{\perp}$ in $A_n$ for some $n$. Since $A_n$ is coherent, 
$a \perp b$ in $\Ev(A_n)$, whence, $a \perp b$ in $\Ev(A)$. Thus, $A$ is coherent. $\Box$


{\bf Lemma C4:} {\em With notation as above, 
$(A^{n})$ is an inductive system of models, and $A^{c} = \bigcup A^{n}$.} 

{\em Proof:} If $\u, \v \in X_n$ and $\u \perp \v$ in $X^{\ast}$, 
then $\u = \u_o x \u_1$, $\v = \u_o y \v_1$ with $x, y \in X(A)$ and $x \perp y$. Thus, 
$\u \perp \v$ in $X_n$. $\bigcup \M(A_n)$ is inductive, and contains 
$\M(A)$, so $\M(A^c) \subseteq \bigcup \M(\for{A}^{n})$. Conversely, 
every inductive set contains every $\M_n$, so 
$\bigcup \M(\for{A}^n) = \M(A^c)$.  It is straightforward that 
$\Omega(A^c) = \Omega(\bigcup \for{A}^{n})$. $\Box$ 

We now assemble these pieces to prove Proposition 2:
that if $A$ is unital, strongly unital, regular, or algebraic, so is $A^{c}$, and if $A$ is  unital and has a strictly positive state, then $A^{c}$ is strongly 
unital. Indeed, by Lemma B2, every $\for{A}^{n}$ has the indicated property; hence, by Lemmas B3 and B4, so does $A^c = \bigcup \for{A}^{n}$. $\Box$ \\

{\bf Morphisms on the Compound Test Space} 

Test-preserving morphisms play well with 
forward products (and, more generally, with Dacey sums). 

{\bf Lemma C5:} {\em Let $\phi$ be a 
test-preserving morphism $A \rightarrow A$. Then 
the unique extension of $\phi : X(A) \rightarrow X(A)$ to 
a mapping $\phi^{c} : X(A)^{\ast} \rightarrow X(A)^{\ast}$ 
with $\phi^{c}(\a x) = \phi^{c}(\a)\phi(x)$ 
for $\a \in X(A)^{\ast}$ and $x \in X(A)$, is a 
test-preserving morphism $A^{c} \rightarrow A^{c}$. }
{\em Proof:} Arguing inductively, one can check that $\phi^c$ is locally injective and 
test-preserving. $\Box$ \\

{\large \bf Appendix D: Some Gratuitously Gory Details} 


We want to show that 
$\#$ is a monad in $\Mod$. So we need to verify 
that the following commute. 
\[
\begin{tikzcd}
A^{\# \# \#} \arrow[d, "\mu_{A^{\#}}" left] \arrow[rr, "\#(\mu_{A})"] & & A^{\# \#} \arrow[d, "\mu_A"] 
\arrow[d, "\mu_A"] \\
A^{\# \#}  \arrow[rr, "\mu_{A}"] & & A^{\#}\\
\end{tikzcd}
\hspace{.2in} 
\begin{tikzcd}
A^{\#} \arrow[drr, "\id_{A}" below] \arrow[rr, "\#(\eta_A)"] & & A^{\# \#} \arrow[d, "\mu_A"] & & A^{\#} \arrow[ll, "\eta_{A^{\#}}" above] \arrow[dll, "\id_{A}"]\\
 & & A^{\#} & & \\
 & & & & 
\end{tikzcd}
\]
For the first of these, let $a \in X(A^{\# \# \#})$: 
$a$ is a sub-partition of a sub-partition of a test 
in $A$. That is, $a = \{ a_{i} | i \in I\}$ 
where each set $a_i$ is itself a collection 
of pairwise-disjoint, non-empty events $a_i = \{a_{i,j} | j \in J_{i}\}$, with 
where $\bigcup_{j \in I_j} a_{i,j} = \bigcup a_i$  
an event of $A$, $\{ \bigcup a_i | i \in I\}$ again 
pairwise disjoint, and $\cup \cup a = \bigcup_{i \in I} \bigcup a_i$ again an event.  Now 
\[\mu_{A^{\#}}(a) = \cup a = \{ a_{i,j} | i \in I, j \in J_i\}\]
and $\mu_{A}$ takes this to $\bigcup \bigcup a 
= \{ x | \exists i \in I, j \in J_{i}  x \in a_{i,j}\}$. 
On the other hand, going across the top of the diagram, 
$\#(\mu_{A})$ takes $a = \{ a_{i} | i \in I\}$ 
to $\{ \mu_{A}(a_i) | i \in I\} = \{ \cup a_i | i \in I\}$, 
and $\mu_{A}$ takes this to $\cup \cup a$. 

For the second diagram, we start with $a \in X(A^{\#})$, i.e., $a \subseteq X(A)$ a non-empty event. 
\[\#(\eta_{A})(a) = \{ \eta_{A}(x) | x \in a\} 
= \{ \{x\} | x \in a\},\]
and $\mu_{A}$ takes this to $\bigcup \{\{x\} | x \in A\} 
= a$. This verifies the commutativity of the left-hand 
triangle. For the right-hand side, again start 
with $a \in X(A^{\#})$, and apply $\eta_{A^{\#}}$: 
this gives us $\{a\}$, and $\mu_{A}(\{a\}) = \bigcup\{a\} 
= a$. $\Box$ 

{\bf Proof of Lemma 10} That $\ell_{A} : A^{\#c} \rightarrow A^{c\#}$ is a test-preserving embedding is straightforward. 
To check naturality, suppose $\phi : A \rightarrow B$ is 
a morphism in $\Prob$. We need to show that 
$\ell_{B} \circ \phi^{\# c} = \phi^{\# c} \circ \ell_{A}$. 
Let $(a_1,...,a_n) \in (A^{\#})^c$ with $a_i \in \Ev(A)$. Then we have 
\begin{eqnarray*}
(\ell_{B} \circ \phi^{\# c})(a_1,...,a_n) 
& = & \ell_{B}(\phi(a_1),...,\phi(a_n)) \\
& = & \phi(a_1) \times \cdots \times \phi(a_n) \\
& = & \phi^{c\#}(a_1 \times \cdots a_n)\\
& = & \phi^{c \#} \ell_{A}(a_1,...,a_n) \ \ \Box 
\end{eqnarray*}

{\bf Proof of Lemma 11}
We need to verify that 
\[\ell_{A}: a_1, \cdots, a_n \in {\cal P}(X)^{n} \mapsto a_1 \cdots a_n \subseteq X^{\ast}\]
satisfies the conditions for a distributive law 
linking $\#$ and $(\cdot )^{\#}$. One relatively 
painless way to do this is to exploit \cite[pp. 314-317, Propositions 
2 and 3]{Barr-Wells},  by showing that $\#$ has a lifting to 
the category $\Prob^{c}$ of $(\cdot)^c$-algebras. This is 
fairly straightforward, since the coarsening of a sequential 
model is still a sequential model under element-wise 
multiplication of events. 
\tempout{
  A lifting of a monoid 
$T$ to $\Cat^{S}$, where $S$ is another monoid on $\Cat$, 
is a monad $(T^{S}, \eta^{S},\mu^{S})$ on $\Cat^{S}$ with 
\begin{itemize} 
\item[(a)] $U^{S}  \circ T^{S} = T \circ U^{S}$; 
\item[(b)] $U^{S} \eta^{S} = \eta U$, and both are 
natural transformations $U^{S} \rightarrow U^{S} \circ T^{S}$; 
\item[(c)] $U^{S} \mu^{S} = \mu U^{S}$, and both 
are natural transformations $U^{S} \circ (T^{S})^2 \rightarrow U^{S} \circ T^{S}$. 
\end{itemize} 
where $U^{S} : \Cat^{S} \rightarrow \Cat$ is the 
underlying (forgetful) functor taking an $S$-algebra 
$(A,\alpha)$ to $A$, and an $S$-algebra homomorphism 
to itself, regarded simply as a morphism in $\Cat$. 
In our case, where $S = ( ~\cdot~)^c$, this means we 
want to define a monad --- let me simply call it 
$(T,\eta,\mu)$ --- on the category of sequential models and 
morphisms, such that $T(A) = A^{\#}$, 
$\eta(x) = \{x\}$, and $\mu$...}
However, we'll take the low road and verify the necessary 
identities directly. 

We need a way of handling some moderately nasty book-keeping: we will be dealing with strings in $X^{\ast}$, subsets of such strings, strings of subsets, 
subsets of {\em these}, and so on. As a visual aid, 
I'm going to write a string as an ordered $n$-tuple, 
that is, $(x_1,...,x_n)$ will stand for the string 
$x_1 \cdots x_n$ in $X^{\ast}$. 
If $(a_1,...,a_n)$ is a string consisting of 
subsets of $X^{\ast}$, then I'll write 
$a_1 \times \cdots \times a_n$ for the set 
of {\em strings} $(x_1,...,x_n)$ with $x_i \in a_i$. 
That is, $a_1 \times \cdots \times a_n \subseteq X^{\ast}$. 
This will help us keep track of things, but carries 
some potential for confusion: again, 
$a_1 \times \cdots \times a_{n}$ is a set of {\em strings}, 
not literally a set of ordered $n$-tuples built 
up inductively from ordered pairs\footnote{Unless, of course, that's how we want to view strings. My own preference 
is to treat a string as a partial function 
$x : \N \rightarrow X$ with domain $\{1,...,n\}$. }  
In particular --- and this will be important below --- if we have e.g., if $a_1, a_2, a_3, a_4 \subseteq X^{\ast}$, $(a_1 \times a_2) \times (a_{3} \times a_{4}) 
= a_1 \times a_2 \times a_3 \times a_4$, where 
the identity is meant literally. 

Now let's dive in. We have four diagrams to check. 

(I) We need to check that 
\[
\begin{tikzcd}
A^{\#\#c} \arrow[d, "(\mu^{\#}_{A})^{c}" left] \arrow[r, " \ell_{A^{\#}}"] 
& 
A^{\# c \#} \arrow[r, "\# \ell_{A}"]
& A^{c \# \#} \arrow[d, "\mu^{\#}_{A^{c}}"]  \\
A^{\# c} \arrow[rr, "\ell_{A}"]  & & A^{c \#} \\
\end{tikzcd}
\]
commutes. A typical element of $A^{\# \# c}$ is a string $a_1...a_n$ where $a_i \in X(A^{\# \#})$. 
This means that each $a_i$ is a non-empty subset 
of some test ${E_{i}}^{\#} \in \M(A^{\#})$, that is, 
some partition $\{a_{i,j}\}$ of a test $E_i \in \M(A)$. 
Thus, $a_{i}$ is a collection, say $\{a_{i,j} | j = 1,...,n_i\}$, of disjoint, non-empty events, with 
union $\bigcup a_i \in \Ev(A)$. The mapping 
$\ell_{A^{\#}}$ takes $a_1,...,a_n$ to 
its set-wise product $a_1 \times \cdots \times a_n \subseteq (X(A^{\#}))^{\ast}$, which consists of all 
strings $(a_{1,j_1}, a_{2,j_2}, \cdots ,  a_{n,j_n}) \in (X(A)^{\#})^{\ast}$ 
with $a_{i,j_i} \in a_i$. The mapping 
$\#\ell_A$ now takes this set to the set of all 
images $\ell_{A}(a_{1,j_1}, \ldots , a_{n,j_n}) 
= a_{1,j_1} \times \cdots \times a_{n,j_n}$. That is, 
\[\#\ell_{A} (a_{1,j_1}, \ldots a_{n,j_n}) 
= \{ a_{1,j_1} \times \cdots \times a_{n,j_n} | 
i = 1,...,n, j_i = 1,...,n_i\}.\]
Finally, $\mu^{\#}_{A^c}$ takes this to 
\begin{equation} 
\bigcup \{ a_{1,j_1} \times \cdots \times a_{n,j_n} | a_{i,j_i} \in a_{i}\} 
= \{ (x_1,...,x_n) | x_i \in a_{i,j_i} \forall i, j_i\}.
\end{equation}
Traveling down the left hand side of the diagram, 
$(\mu^{\#}_{A})^{c}$ takes our original string 
$(a_1,...,a_n)$ to the string $(\bigcup a_1,...,\bigcup a_n)$ in $X(A^{\#})^{\ast}$. Appling $\ell_{A}$, 
this is mapped to 
\[\cup a_1 \times \cdots \cup a_n 
= \{ (x_1,...,x_n) | x_i \in \cup a_i \forall i \} 
= \{ (x_1,...,x_n) | \forall i \exists j_i x_i \in a_{i,j_i}\}.\]
This is the same set as the one in (\theequation). 

(II) Next, we verify that 
\[
\begin{tikzcd}
A^{\#cc} \arrow[d, "\mu^{c}_{A^{\#}}" left] \arrow[r, " (\ell_{A})^{c}"] 
& 
A^{c \# c} \arrow[r, "\ell_{A^{c}}"]
& A^{c c \#} \arrow[d, "\#(\mu^{c}_{A})"]  \\
A^{\# c} \arrow[rr, "\ell_{A}"]  & & A^{c \#} \\
\end{tikzcd}
\]
commutes. This is a bit easier. An outcome 
in $X(A^{\# cc})$ is a string of strings of events of $A$. 
That is, $(a_1,...,a_n)$ where $a_i = (a_{i,1},...,a_{i,k_i})$, with $a_{i,j} \in \Ev(A)$ for all $i,j$. 
$(\ell_{A})^{c}$ 
takes this to the string of sets 
\[(a_{1,1} \times \cdots \times a_{1,k_1}), 
\cdots , (a_{n,1} \times \cdots \times a_{n,k_n}).\]
Under $\ell_{A^c}$, this last goes to the set of strings 
\begin{equation} 
(a_{1,1} \times \cdots \times a_{1,k_1}) 
\times \cdots \times (a_{n,1} \times \cdots \times a_{n,k_n}).
\end{equation} 
On the other hand, if we first apply $\mu^{c}_{A^{\#}}$ 
to $(a_1,...,a_n)$, we get the long string of sets 
\[(a_{1,1},..., a_{n,k_{n}}).\]
This is mapped by $\ell_{A}$ to the product set 
\[a_{1,1} \times \cdots \times a_{n,k_{n}}.\]
{\em Since we are dealing with strings and 
setwise products in $X^{\ast}$ and $\P(X^{\ast})$}, 
this is the same (not just canonically isomorphic to) 
the set in (\theequation). 

(III) We need to show that 
\[
\begin{tikzcd}
 & A^{\#}  \arrow[dl, "\eta^{c}_{A^{\#}}" above left] \arrow[dr, "\#(\eta^{c}_{A})" above right] & \\ 
A^{\# c} \arrow[rr, "\ell_{A}"] & & A^{c\#} \\
& & 
\end{tikzcd}
\]
commutes. An element of $X(A^{\#})$ is an event 
$a \in \Ev(A)$. $(\eta^{c}_{A^{\#}})(a) = a$, understood 
as a length-one string in $X(A^{\#})^{c}$. 
Now $\ell_{A}$ takes this to $a$, understood as an 
event in $\Ev(A^{c})$ understood as a set of length-one
strings in $X^{\ast}$. But $\#(\eta^{c}_{A})$ takes 
$a$ to $\{ \eta^{c}(x) | x \in a\} = a$, the same 
set of length-one strings.  

(IV) Finally, consider 
\[
\begin{tikzcd}
 & A^{c}  \arrow[dl, "(\eta_{A})^{c}" above left] \arrow[dr, "\eta^{\#}_{A^{c}}" above right] & \\ 
A^{\# c} \arrow[rr, "\ell_{A}"] & & A^{c\#} \\
& & 
\end{tikzcd}
\]
An element of $X(A^c)$ is a string $(x_1,...,x_n) \in X(A)^{c}$; $c(\eta^{\#}_{A})$ takes this to 
$(\eta^{\#}_{A}(x_1),...,\eta^{\#}_{A}(x_n)) 
= (\{x_1\},...,\{x_n\})$, and $\ell_{A}$ takes 
{\em this} to $\{x_1\} \times \cdots \times \{x_n\} 
= \{(x_1,..,x_n)\} = \eta^{\#}_{A^c}(x_1,...,x_n)$. 

This finishes the job. $\Box$ \\


\end{document}